\documentclass{article}

\usepackage{arxiv}

\usepackage[utf8]{inputenc} 
\usepackage[T1]{fontenc}    
\usepackage{hyperref}       
\usepackage{url}            
\usepackage{booktabs}       
\usepackage{multirow}       
\usepackage{amsfonts}       
\usepackage{nicefrac}       
\usepackage{microtype}      
\usepackage{graphicx}
\usepackage[version=4]{mhchem}
\usepackage[
    style=numeric-comp,
    sorting=none,
    maxbibnames=5,
    minbibnames=3
]{biblatex}
\usepackage{doi}
\usepackage{svg}
\usepackage{amsmath,amssymb}
\usepackage[capitalize]{cleveref}
\usepackage{xspace}
\usepackage{tabularx}
\usepackage{multicol}
\usepackage{setspace}
\usepackage[most]{tcolorbox}
\usepackage{xcolor}
\usepackage{listings}
\usepackage{ragged2e}
\usepackage{float}
\usepackage[linesnumbered,ruled,vlined]{algorithm2e}

\SetAlgorithmName{Workflow}{Workflow}{List of Workflows}
\SetKwInput{KwInput}{Input}
\SetKwInput{KwOutput}{Output}
\SetKw{KwRet}{return}
\SetKwFunction{UserMsg}{UserMsg}
\SetKwFunction{AIMsg}{AIMsg}
\SetKwFunction{LLM}{LLM}
\SetKwFunction{LLMJson}{LLM\_Json}
\SetKwData{History}{Messages}
\SetKwFunction{Format}{Format}
\SetKwFunction{GenRead}{GenReadPrompt}
\SetKwFunction{GenRetry}{GenRetryPrompt}
\SetKwFunction{GenMatch}{GenMatchPrompt}
\SetKwFunction{GenRefine}{GenRefinePrompt}
\SetKwFunction{GetUnmatched}{GetUnmatchedCodes}
\SetKwFunction{Join}{Join}
\SetKwFunction{GenDetect}{GenDetectPrompt}
\SetKwFunction{GetMatched}{GetMatchedValues}
\SetKwFunction{Keys}{Keys}

\SetCommentSty{mycommfont}

\lstdefinelanguage{prompt}{
    basicstyle=\small\ttfamily\linespread{1.1}\selectfont,
    breaklines=true,
    extendedchars=true,
    columns=fullflexible,
    keepspaces=true,
    frame=none,
    literate={⋅}{{$\cdot$}}1
             {Å}{{\AA}}1
             {°}{{$^{\circ}$}}1
             {³}{{$^3$}}1
             {•}{{$\bullet$}}1
             {…}{{...}}1
             {"}{{"}}1
             {"}{{"}}1
             {'}{{'}}1
             {'}{{'}}1
             {–}{{--}}1
             {—}{{---}}1,
    moredelim=**[is][\color{blue}\bfseries]{|}{|}
}

\newtcolorbox{promptbox}[1][]{
    enhanced,
    breakable,
    title=\textbf{System Prompt Used for Extraction},
    colback=gray!5,
    colframe=black!70,
    width=\linewidth,
    boxrule=0.5pt,
    arc=2mm,
    fontupper=\small,
    #1
}

\title{LitMOF: An LLM Multi--Agent for Literature--Validated Metal--Organic Frameworks Database Correction and Expansion}

\date{}

\author{
\begin{tabular}{ccc}
    \textbf{Honghui Kim}\textsuperscript{1} &
    \textbf{Dohoon Kim}\textsuperscript{1} &
    \textbf{Jihan Kim}\textsuperscript{1}\thanks{Corresponding author} \\
    \texttt{ildmb96@kaist.ac.kr} &
    \texttt{dhoonkim97@kaist.ac.kr} &
    \texttt{jihankim@kaist.ac.kr}
\end{tabular}
\\[2.5em]
\parbox{0.9\textwidth}{\centering
\textsuperscript{1}Department of Chemical and Biomolecular Engineering,
Korea Advanced Institute of Science and Technology, Daejeon, Republic of Korea
}
}

\renewcommand{\headeright}{}
\renewcommand{\undertitle}{}
\renewcommand{\shorttitle}{}

\hypersetup{
pdftitle={LitMOF: An LLM Multi--Agent for Literature--Validated Metal--Organic Frameworks Database Correction and Expansion},
pdfsubject={},
pdfauthor={Honghui Kim, Jihan Kim},
pdfkeywords={},
colorlinks=false,
pdfborder={0 0 1},
linkbordercolor={0 0 1},
citebordercolor={0 0.6 0},
urlbordercolor={0 0.6 0.6},
filebordercolor={1 0 0},
}

\begin{document}

\begin{refsection}

\maketitle

\begin{abstract}
Metal--organic framework (MOF) databases have grown rapidly through experimental deposition and large-scale literature extraction, but recent analyses show that nearly half of their entries contain substantial structural errors.
These inaccuracies propagate through high-throughput screening and machine-learning workflows, limiting the reliability of data-driven MOF discovery.
Correcting such errors is exceptionally difficult because true repairs require integrating crystallographic files, synthesis descriptions, and contextual evidence scattered across the literature.
Here we introduce LitMOF, a large language model--driven multi-agent framework that validates crystallographic information directly from the original literature and cross-validates it with database entries to repair structural errors.
Applying LitMOF to the experimental MOF database (the CSD MOF Subset), we constructed LitMOF-DB, a curated set of 189,567 computation-ready structures, including the successful repair of 9,277 invalid entries, which accounts for 69.1\% of the CSD-derived not-computation-ready MOFs in the latest CoRE MOF DB.
Additionally, the system uncovered 8,771 synthesized but undeposited MOFs absent from existing resources, substantially expanding the known experimental design space.
Using direct air capture screening as a case study, we demonstrate that structural errors severely distort predicted adsorption energies and \ce{CO2}/\ce{H2O} selectivity, leading to systematic misranking of materials, false positives, and the omission of high-performance candidates.
This work establishes a scalable pathway toward self-correcting scientific databases and a generalizable approach for LLM-driven curation in materials science.
\end{abstract}


\section{\label{sec:intro}Introduction}
Metal--organic frameworks (MOFs) constitute one of the most extensively studied classes of porous materials, with high structural diversity arising from the modular combination of metal nodes, organic linkers, and network topologies\cite{yaghi2003reticular}.
This diversity has motivated creation of large curated databases, including the CoRE MOF databases\cite{chung2014computation, chung2019advances, zhao2025core}, the CSD MOF subset\cite{allen2002cambridge, moghadam2017development}, QMOF database\cite{rosen2021machine, rosen2022high}, and MOSAEC-DB\cite{gibaldi2025mosaec}, which serve as the foundation for high-throughput simulations\cite{colon2014high, lee2021computational, kwon2019computer}, machine learning based property prediction\cite{jablonka2021using, burner2020high, moghadam2019structure, batra2020prediction, han2024machine}, and accelerated materials discovery \cite{kang2023multi, terrones2024metal, park2024inverse}.
The accuracy and reliability of these databases are therefore central to all computational and data-driven MOF research.

However, recent work by White et al. revealed that structural errors are widespread across computation-ready MOF databases\cite{white2025high}.
Using the metal oxidation state algorithm (MOSAEC), they reported that 51\% of entries (>1.9 million structures) across 14 leading MOF databases violate basic chemical valence principles, and that 52\% of top candidates in recent high-throughput screening studies are chemically invalid.
Although several studies had previously noted the presence of structural inconsistencies\cite{velioglu2020revealing, daglar2021effect, chen2020identifying, gibaldi2022healed}, the scale of the issue had been significantly underestimated.

Current approaches to mitigate these errors, including the rule-based sanity checks\cite{Jablonka_mofchecker_2023}, improved solvent removal workflows\cite{zhao2025core, gibaldi2024incorporation}, and MOSAEC curation pipeline\cite{white2025high}, share a fundamental limitation in that they were designed to identify invalid structures, but not to repair them.
Because these methods rely on fixed chemical rules and heuristic criteria, they cannot determine the correct structure when inconsistencies arise.
As a result, large fractions of experimentally reported MOFs remain unusable for simulation.
For example, in the latest CoRE MOF DB\cite{zhao2025core, zhao_2025_15055758}, 7635 entries are designated as computation-ready, while 11,764 additional structures derived from the CSD fail to meet computational criteria after free-solvent removal\cite{zhao2025core}.
This reflects the inherent difficulty of processing experimental CIFs into computation-ready CIFs.
Moreover, MOFs reported in publications but not deposited as complete CIFs are excluded entirely, limiting coverage of the experimentally known space.
Repairing these errors requires reconciling information scattered across multiple sources (e.g. original publications, crystallographic repositories, database records).
But to date, performing this reconciliation manually for hundreds of thousands of structures has remained impossible.

To remedy this, we introduce LitMOF, the first LLM-driven multi-agent framework that can both detect and repair structural errors in experimental MOF entries by validating them against primary literature.
LitMOF automatically retrieves evidence from (1) the literature (2) existing databases (CSD, CoRE MOF DB, MOSAEC-DB), and (3) crystallographic files, and cross-validates these resources to reconstruct chemically valid structures as originally synthesized and reported.

Applying this framework to the CSD MOF subset,
we constructed LitMOF-DB, comprising 120,419 unique experimental MOFs.
For each MOF, one or more computation-ready structures are generated by systematically varying the degree of non-framework component removal, resulting in a total of 189,567 validated structures.
Notably, LitMOF-DB successfully recovers 9,277 previously non-computation-ready CoRE MOF DB structures (originating from the CSD), which accounts for 69.1\% of all such invalid entries, demonstrating the ability of the framework to transform previously unusable structures into computation-ready form.
Moreover, during the curation process, we identified 8,771 synthesized but undeposited MOFs, present in the literature but absent from the CSD, further expanding the known experimental space.
To assess the downstream consequences of structural errors, we further use direct air capture screening as a representative application case and show that erroneous structures systematically distort adsorption thermodynamics and materials ranking.

Together, LitMOF and LitMOF-DB substantially improve the structural fidelity of MOF databases and provide a scalable pathway for incorporating missing experimental knowledge into computation-ready resources.
This framework illustrates how LLM-powered agents can transform materials curation and lays the foundation for dynamic, self-correcting databases across diverse materials classes.

\section{\label{sec:results}Results}
\subsection{LitMOF Agent Overview}

\begin{figure}[!ht]
	\centering
	\includegraphics[width=\textwidth]{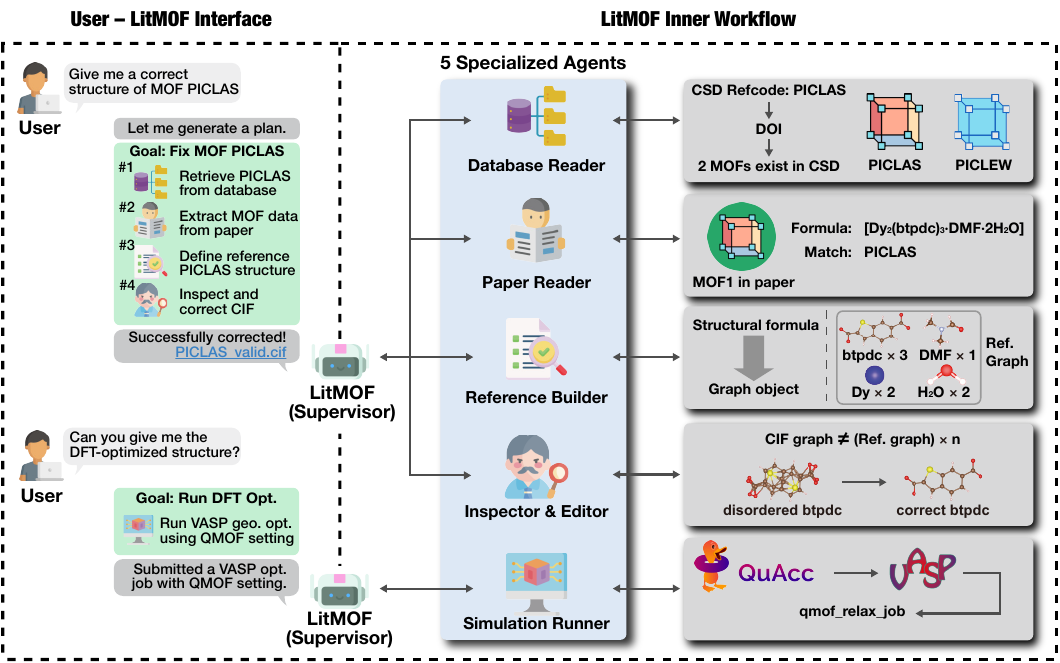}
	\caption{
    Schematic illustration of how LitMOF multi-agent system interacts with a user and generates responses.
    LitMOF consists of a Supervisor and five specialized agents, and the Supervisor interprets the user query and dispatches tasks to the appropriate agents.
    For the PICLAS example, LitMOF retrieves database records, extracts information from the publication, constructs a reference graph, and corrects structural errors in the CIF.
    LitMOF can also execute follow-up tasks, such as DFT geometry optimization, via the Simulation Runner.
    }
	\label{fig:overview}
\end{figure}

Constructing large-scale databases requires workflows that can integrate information from diverse sources and correct structural inconsistencies automatically.
Traditional automation has relied on hand-coded rules that operate only on structured data, whereas human experts routinely combine structured information with unstructured descriptions in publications to determine the correct MOF structure.
To emulate this human-like reasoning, we developed LitMOF, a multi-agent system that leverages LLMs to automate the inspection and repair of structures.

The LitMOF system consists of a Supervisor together with five specialized agents: Database Reader, Paper Reader, Reference Builder, Inspector \& Editor, and Simulation Runner.
The Database and Paper Reader retrieve structural information of MOFs from databases and publications.
The Reference Builder constructs the expected structural motif from these sources, and the Inspector \& Editor identifies and corrects inconsistencies and errors in the CIF.
Finally, the Simulation Runner performs optional computational simulations on corrected structures.
The Supervisor orchestrates all agents, maintains the execution plan, and interacts directly with the user.

\Cref{fig:overview} illustrates the overall workflow.
When a user requests the corrected structure of a MOF (e.g. CSD Refcode PICLAS), the Supervisor generates a plan that coordinates the four relevant agents:  retrieving database records, extracting information from the associated publication, constructing the reference graph, and applying structural corrections.
The validated structure is then returned to the user.
In a second workflow, a user may request a DFT-optimized structure;  in this case, the Supervisor includes the Simulation Runner, which prepares input files using the corrected CIF and submits a geometry optimization job.
These examples highlight how LitMOF interprets user queries, decomposes the query into coordinated multi-agent tasks, and delivers corrected or computation-ready MOF structures.
The detailed logic and plan-execution behavior of each agent are described in \Cref{sec:spec_agents}.

\subsection{Plan-and-Execute Agent Architecture}

\begin{figure}[!ht]
	\centering
	\includegraphics[width=\textwidth]{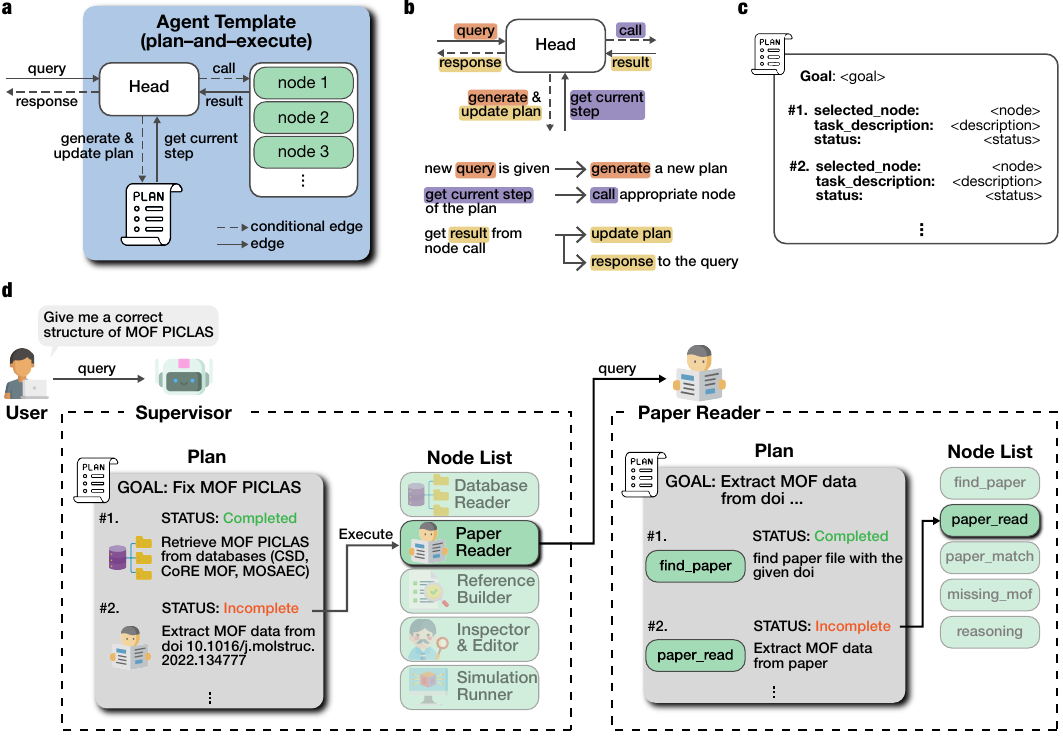}
	\caption{
    \textbf{a}, Unified agent template comprising an LLM-driven head module and a set of nodes, each representing either another agent call or an LLM/tool operation.
    \textbf{b}, Decision process of the head module, which interprets the query, generates or updates a plan, selects the next node, and determines termination.
    \textbf{c}, Structure of an agent plan, represented as a overall goal and an ordered list of nodes with associated descriptions and execution statuses.
    \textbf{d}, Hierarchical plan-and-execute behaviour illustrated using the PICLAS correction workflow, where the Supervisor's high-level plan expands into finer-grained plans executed by specialized agents.
    }
	\label{fig:plan_execute}
\end{figure}

The LitMOF system adopts a plan-and-execute agent architecture\cite{wang2023plan} in which every agent including the Supervisor and the five specialized agents follows a unified execution template (\Cref{fig:plan_execute}a).
Each agent is composed of a head module and a list of nodes, where a node may correspond to another agent responsible for a specific subtask or to an LLM/tool call that executes a well-defined function.
The head module receives a query, generates or updates a plan, and determines the next node to execute.
Powered by an LLM, the head module selects the next action based on the agent's current state and the intermediate outputs produced during execution.
\Cref{fig:plan_execute}b illustrates representative scenarios in which the head module makes decisions.
When an agent receives an external query, the head first interprets the query and generates an initial plan.
If a plan already exists, the head module retrieves the current step from the plan and invokes the corresponding node.
After the node is executed, the head receives the result, updates the plan accordingly, and determines the subsequent action.
Throughout this process, the head module also decides when the plan is complete and when the agent should terminate and return a response.
The plan consists of an overall goal to be achieved within the agent and an ordered list of subtasks, each represented by a node with its own description and execution status (see \Cref{fig:plan_execute}c).
During execution, the head sequentially selects the next node, invokes the associated function, records the result, and updates the plan status. This mechanism enables agents to operate autonomously while maintaining consistency across the entire system.

Because every agent follows the plan-and-execute style, the LitMOF system as a whole forms a hierarchical plan-and-execute architecture.
\Cref{fig:plan_execute}d illustrates the hierarchical structure of the plan-and-execute architecture using the workflow for correcting the MOF PICLAS as an example.
When the Supervisor receives the user's query, it generates a plan composed of coarse-grained tasks such as retrieving database records or extracting information from the publication.
Each of these tasks is delegated to a specialized agent, which in turn generates its own plan consisting of finer-grained nodes tailored to its domain-specific function.
For instance, the Paper Reader expands the Supervisor's "extract MOF data from paper" task into subtasks such as locating the paper file, reading the paper, matching components, and resolving missing information.
As each agent executes its plan and returns its results, the Supervisor integrates the outputs and advances the overall workflow.
This nested structure enables LitMOF to decompose complex queries into coordinated, interpretable, and reusable multi-agent plans.

\subsection{\label{sec:spec_agents}Specialized Agents in LitMOF}

The Database Reader retrieves structural metadata from the CSD, CoRE MOF DB, and MOSAEC-DB.
For CSD entries, it uses the CSD Python API to obtain information such as the DOI, chemical name, chemical formula, and lattice parameters.
The full list of CSD metadata fields collected by the Database Reader is summarized in \Cref{tab:csd_fields}.
For the CoRE MOF DB and MOSAEC-DB, the agent checks whether a CIF file corresponding to a given Refcode is available.
In the PICLAS example (\Cref{fig:overview}), no CIF associated with the Refcode PICLAS exists in either CoRE MOF DB or MOSAEC-DB, whereas CSD contains a valid entry.
From the CSD, the Database Reader first retrieves the DOI linked to PICLAS and identifies all MOF structures reported under that DOI.
For PICLAS's DOI, two structures (PICLAS and PICLEW) are present, and metadata are collected for both entries to support downstream tasks in the Paper Reader.

\begin{figure}[!ht]
	\centering
	\includegraphics[width=0.6\textwidth]{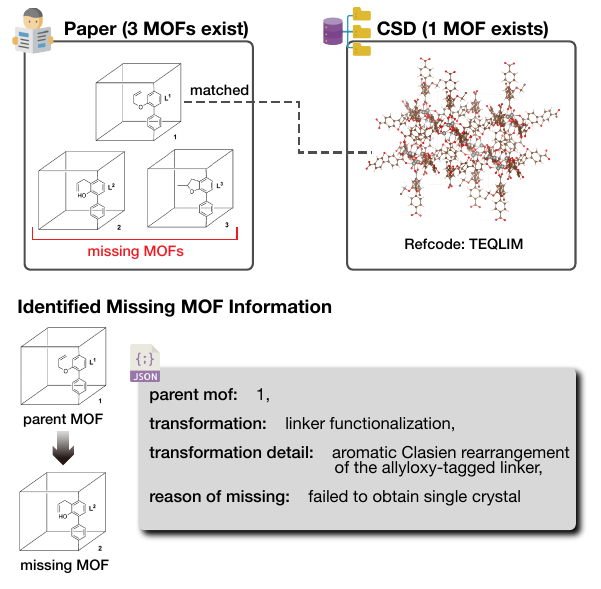}
	\caption{
    Example of a missing MOF case (refcode: TEQLIM).
    Missing MOFs refer to structures that were synthesized and characterized in the literature but were not deposited as CIF files in the CSD.
    This example contains two such missing MOFs.
    For each missing MOF, LitMOF identifies the parent MOF (the most structurally similar MOF available in the CSD), the transformation required to obtain the missing MOF from its parent, and the reason the CIF is missing when explicitly provided in the paper.
    }
	\label{fig:missing_mof}
\end{figure}

The Paper Reader identifies publications relevant to a given MOF and extracts structural and chemical information from the text.
Rather than employing retrieval-augmented generation (RAG), the Paper Reader performs full-document inference on the entire parsed text of each publication, which we found to be more reliable for MOF-specific information extraction in a single document (see \Cref{sec:rag_vs_full}).
In addition, the Paper Reader employs dynamic prompting to iteratively refine extraction, matching, and detection tasks based on intermediate results, as described in \Cref{sec:dynamic_prompt}.
Its core functions include 1) locating paper file and parse the text, 2) extracting MOF specific metadata, 3) matching extracted information to CSD Refcodes, and 4) identifying missing MOFs that are synthesized but not deposited in CSD.
List of extractable metadata from a paper is summarized in \Cref{tab:paper_reader_descrip}.
In the PICLAS example, the Paper Reader first locates the publication using the DOI retrieved by the Database Reader.
The identified file is then parsed to obtain cleaned text in either plain text or markdown format (a list of supported file formats and publishers is provided in \Cref{tab:file_ext_support}).
Using this parsed text, the Paper Reader extracts metadata for the two MOFs reported in the publication, including their structural formula, metal node, organic linker, solvent information, and physicochemical properties.
These extracted MOFs are then matched to the two Refcodes, PICLAS and PICLEW, by comparing metadata obtained from the Database Reader and the Paper Reader.
After matching, the agent optionally checks for MOFs that appear in the publication but do not correspond to any deposited CSD entry.
We refer to these structures as missing MOFs.
For each missing MOF, the agent identifies the most similar MOF among those with matched Refcodes (denoted as the parent MOF) and determines the transformation that converts the parent MOF into the missing MOF (see \Cref{fig:missing_mof}).

The Reference Builder defines the expected reference structure of a MOF using information obtained from both the Database Reader and the Paper Reader.
We define the reference structure as the minimal repeating unit of the MOF, which is typically described by its structural formula.
In the PICLAS example, the structural formula extracted by the Paper Reader is \ce{[Dy2(btpdc)3*DMF*2(H2O)]}, indicating that the minimal repeating unit consists of two \ce{Dy} atoms, three \ce{btpdc} linkers, one \ce{DMF} molecule, and two \ce{H2O} molecules.
After determining this minimal repeating unit, the Reference Builder converts each component into a graph object.
This transformation is supported by a combination of the PubChem API, an IUPAC name parser, and a predefined name to SMILES mapping.
Because the Paper Reader provides expanded names for all abbreviations (for example, \ce{H2btpdc} = benzo[b]thiophene-2,6-dicarboxylic acid), the Reference Builder is able to generate complete structural graphs for all components, resulting in the final reference graph shown in \Cref{fig:overview}.
The reference graph can be further validated and combined into a single graph object if CSD provides a valid chemical diagram of a MOF (see \Cref{fig:ref_graph}).

The Inspector \& Editor evaluates whether a CIF structure is consistent with the reference graph and corrects the structure when discrepancies are found.
The agent first constructs a graph representation of the CIF.
The source of CIF is either by using the CSD Python API or by parsing the provided CIF file.
This CIF graph is then compared with the reference graph at several levels of structural detail to determine whether the structure is valid (see \Cref{fig:comparison_test})
The first comparison assesses whether the species composition of the CIF follows the stoichiometric pattern encoded in the reference structure (see \Cref{fig:comparison_test}a).
This is examined by determining whether all atomic species in the CIF can be related to those in the reference structure through a single scaling factor.
When all elements follow the same scaling factor, the overall species composition is regarded as consistent.
If the species-level scaling pattern is consistent, the agent examines the coordination environment of atoms (see \Cref{fig:comparison_test}b).
The coordination environment is defined by the element type of the atom and the list of neighboring atoms' element types.
Another direct comparison is to find subgraph that matches reference graph in the CIF graph (see \Cref{fig:comparison_test}c).

\begin{figure}[!ht]
	\centering
	\includegraphics[width=\textwidth]{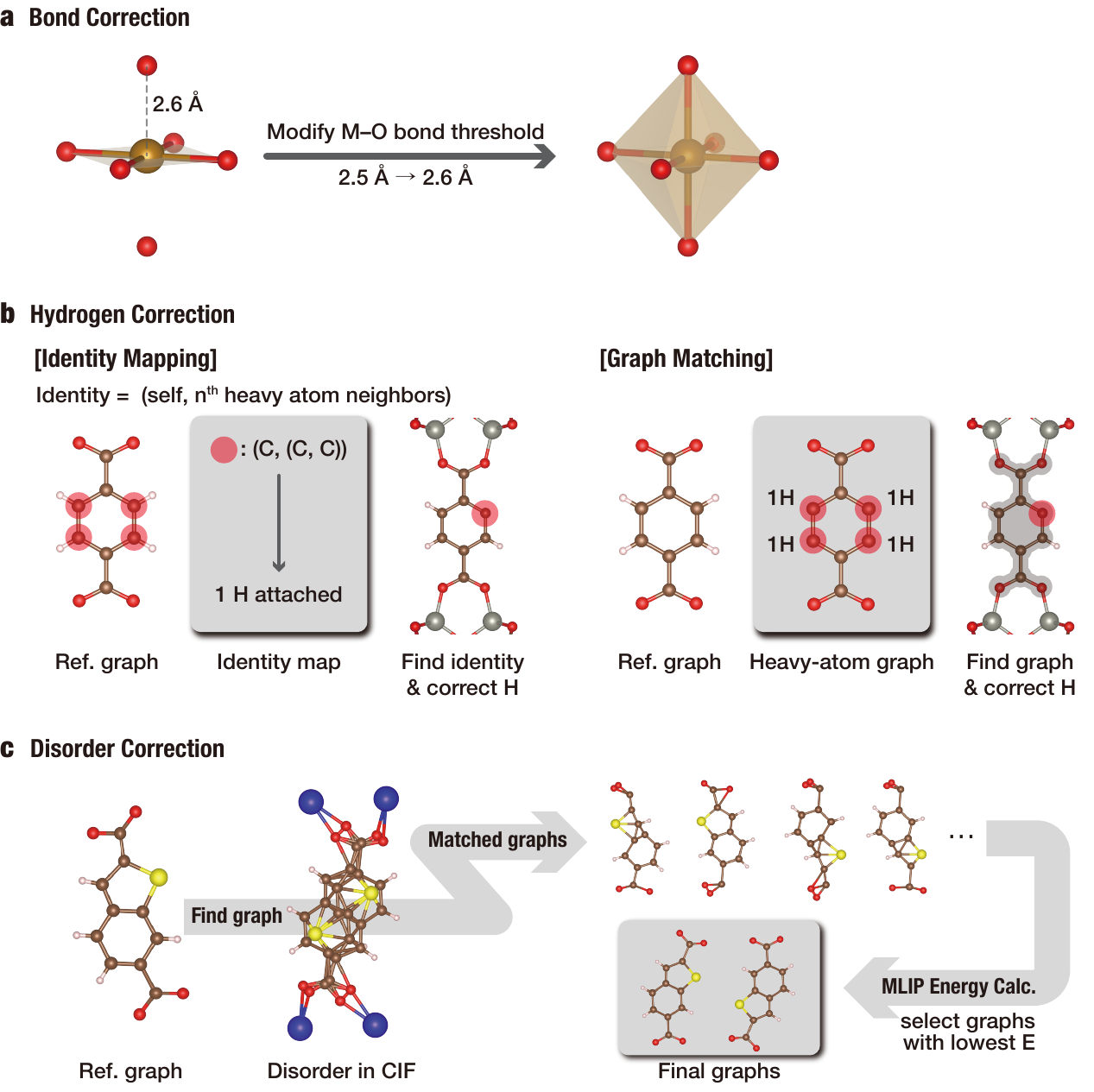}
	\caption{
    Three types of error correction handled by the Inspector \& Editor agent.
    \textbf{a}, Bond errors are corrected by adjusting the distance threshold used to determine bond formation, which adds or removes bonds as needed.
    \textbf{b}, Hydrogen errors are corrected using two complementary methods, identity mapping and graph matching.
    \textbf{c}, Disorder correction resolves duplicated or entangled components into chemically meaningful configurations through graph matching and MLIP-based energy evaluation.
    }
	\label{fig:error_correction}
\end{figure}

Through these comparisons, the Inspector \& Editor identifies three main categories of structural errors, which are bond errors, hydrogen errors, and unresolved disorder.
Bond errors occur when the correct atoms are present, but the inferred bonding network does not match the reference connectivity.
These errors are corrected by adjusting the bond-detection thresholds until the expected network is recovered (see \Cref{fig:error_correction}a).
Hydrogen errors arise when the heavy-atom framework is correct, but hydrogen atoms are misplaced or missing.
To address this issue, the agent identifies heavy-atom sites with incorrect hydrogen counts by comparing coordination-based identity labels or by matching the heavy-atom backbone of the reference graph to the corresponding subgraph in the CIF (see \Cref{fig:error_correction}b).
Once the correct correspondence is established, hydrogen atoms are added or removed as needed.
Unresolved disorder in CIF files results in duplicated fragments or multiple local configurations.
When such disorder is detected, the agent enumerates all candidate structures consistent with the reference graph and evaluates their energies using a machine-learning interatomic potential (MLIP) trained on organic molecules, such as MACE-OFF24\cite{kovacs2025mace} (see \Cref{fig:error_correction}c).
Unphysical candidates typically exhibit higher energies and are removed.

In the PICLAS example, the CIF graph fails the initial species count comparison test against the reference graph.
The CIF graph contains the species counts \{\ce{C}: 176, \ce{H}: 92, \ce{O}: 80, \ce{S}: 16, \ce{Dy}: 8, \ce{N}: 8\}, whereas the reference graph contains \{\ce{C}: 33, \ce{H}: 23, \ce{O}: 15, \ce{S}: 3, \ce{Dy}: 2, \ce{N}: 1\}.
The required repeating factor that aligns the Dy atom count between the two graphs is 4, which yields the scaled reference counts \{\ce{C}: 132, \ce{H}: 92, \ce{O}: 60, \ce{S}: 12, \ce{Dy}: 8, \ce{N}: 4\}.
However, these values still do not match the CIF graph for \ce{C}, \ce{O}, \ce{S}, and \ce{N}, indicating that some components appear in excess relative to the expected minimal repeating unit.
Such discrepancies suggest the presence of duplicated components arising from unresolved disorder in the CIF structure.
This is further confirmed by sub-graph matching. Given that the minimal repeating unit (the reference graph) contains three btpdc linkers and one DMF molecule, a repeating factor of four implies that the CIF should contain twelve btpdc linkers and four DMF molecules.
Instead, the CIF graph contains sixteen btpdc linkers and eight DMF molecules, demonstrating that both components are duplicated beyond their expected multiplicity.
Therefore, the Inspector \& Editor corrects the disorder, as illustrated in \Cref{fig:error_correction}c.
During this process, the entangled btpdc site is disentangled into two possible configurations, resulting in multiple candidate CIF structures.
To identify the most plausible configuration, we compute the energy of each candidate using a universal MLIP trained on materials (e.g., MACE-MPA-0\cite{batatia2025foundation}) and select the one with the lowest energy as the final corrected structure. The implications of having multiple correction solutions are further discussed in the \Cref{sec:multi_sol}.
The MLIP energy is used only to distinguish chemically plausible configurations from implausible ones, a coarse discrimination that does not require highly accurate energies. For the PICLAS linker, for instance, only two of the 1,728 matched candidates lie at the energy minimum, while the rest are separated by relative energies of up to $\sim$34,000~eV from steric clashes (\Cref{fig:piclas_ehisto} and \Cref{fig:piclas_candidates}). When two or more candidates are nearly degenerate in energy, such as orientational configurations of a rotatable linker with dynamic disorder, all of them are valid representations of the disorder, and any one yields a correction-ready structure. The validity of the output therefore does not critically depend on the energy ranking or on the level of theory used to compute it.

Simulation Runner is a specialized agent that can run computational simulation or analysis tool on the corrected CIF.
The agent can run density functional theory (DFT) simulation on the corrected CIF or analyze pore geometry of MOF structure using ZEO++ software\cite{willems2012algorithms}.
DFT runs for MOF structures are powered by Quantum Accelerators package\cite{rosen_2025_17373420}.
The Simulation Runner offers automatic simulation tools so that user can simultaneously correct structure and run simulation on the corrected structure.
As illustrated in \Cref{fig:overview}, the Simulation Runner is called when user wants subsequent computational simulation on the corrected structure.

\subsection{Construction of LitMOF-DB}

\begin{figure}[!ht]
	\centering
	\includegraphics[width=\textwidth]{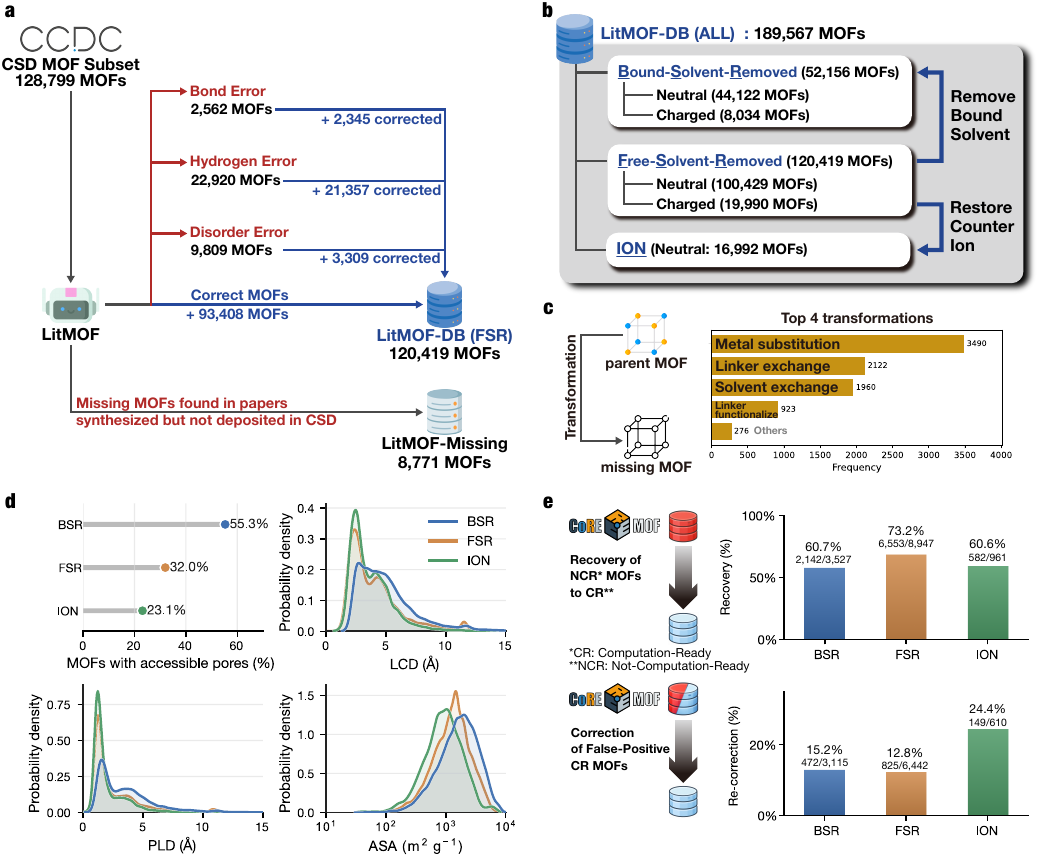}
    \caption{
        \textbf{a}, Results of the database construction using the LitMOF agent, starting from the 128,699 structures of the CSD MOF subset that were processable (100 of the 128,799 subset entries were excluded for CSD-load failure or an empty/unparsable framework).
        We corrected 27,011 MOFs and constructed a curated database of 120,419 experimental MOFs with free solvent removed.
        This database is labeled as LitMOF-DB (FSR).
        During this process, we also identified 8,771 missing MOFs and compiled a separate missing-MOF database.
        \textbf{b}, Expansion of LitMOF-DB (FSR) to other formats: BSR (bound-solvent-removed) and ION (restored charged non-framework components).
        Combining these three database types, LitMOF-DB (ALL) consists of 189,567 MOFs.
        \textbf{c}, The four most common transformations that relate a parent MOF to its corresponding missing MOF.
        \textbf{d}, Pore geometry analysis of LitMOF-DB.
        \textbf{e}, Correction of the latest CoRE MOF DB\cite{zhao2025core, zhao_2025_15055758} during the construction of LitMOF-DB.
        CoRE MOF DB entries derived from the CSD are compared.
        LitMOF-DB (BSR) is compared with the CoRE MOF DB (ASR; all-solvent-removed).
        Because the CoRE MOF DB (ASR) contains duplicate entries of the CoRE MOF DB (FSR) when no bound solvent is present, such duplicates are discarded in this comparison.
    }
	\label{fig:litmof_db_result}
\end{figure}

We applied the automated database curation workflow to the CSD MOF subset, which contains 128,799 experimentally synthesized MOFs.
Of these, 100 entries that failed to load from the CSD or yielded an empty or unparsable framework (for example, structures with no resolvable framework) were excluded, leaving 128,699 structures for processing.
These structures are linked to 53,306 publications, among which we were able to access 36,146 papers from Elsevier, the American Chemical Society, the Royal Society of Chemistry, and Wiley.
The exact counts for each publisher are summarized in the Supporting Information.
Using the LitMOF workflow, we constructed a curated and computation-ready database of 120,419 MOFs with all free, non-coordinating components removed (\Cref{fig:litmof_db_result}a).
This free-solvent-removed database is hereafter referred to as LitMOF-DB (FSR), where FSR denotes free-solvent-removed.
Across the full CSD MOF subset, we identified 2,562 cases of bond errors, 22,920 cases of hydrogen-placement errors, and 9,809 cases of disorder-related errors.
A single MOF may contain more than one error type, but for analysis each MOF was assigned the most severe error type.
LitMOF successfully corrected 2,345 bond-error structures (91.5~\%), 21,357 hydrogen-error structures (93.2~\%), and 3,309 disorder-error structures (33.7~\%).
Structures that could not be fully repaired to a computation-ready state were excluded from LitMOF-DB.
The per-category composition and correction rates of the resulting database are summarized in \Cref{tab:db_snapshot}.
Unlike other types of structural errors, disorder-related errors arise from intrinsic degrees of freedom associated with heavy atoms, including partial occupancies, split atomic positions, and dynamically averaged conformations.
Because these effects prevent the definition of a unique bonding topology, a substantial fraction of disorder-error structures cannot be represented by the predefined reference graph.
In principle, such cases could be addressed by redefining the heavy-atom moieties in the CIF.
One possible approach is to remove the disordered organic linker atoms and insert an idealized linker molecule at chemically plausible positions.
However, this procedure replaces experimentally determined atomic positions with model-derived coordinates and therefore breaks experimental validation.
Since the objective of LitMOF-DB is to provide a computation-ready database of experimentally reported MOFs, such reconstructed structures are deliberately excluded from correction and treated as intrinsically non-repairable within our framework.

To validate the corrected MOFs in LitMOF-DB (FSR), we manually validated 1,000 structures stratified by diagnosed error type (400 hydrogen-error, 200 disorder-error, and 200 bond-error structures, together with 200 structures diagnosed as already correct) by consulting the original publications and relevant databases.
This manual validation resulted in an overall success rate of 98.8~\% (988 correct MOFs) and identified 12 invalid MOFs (\Cref{tab:manual_validation}).
The invalid cases comprise 9 hydrogen-placement errors (out of 400) and 3 disorder cases (out of 200), whereas the bond-error and no-error structures contained no invalid cases.
In some cases, corrections were valid at the level of graph connectivity but resulted in unrealistically short H--H distances due to experimentally constrained heavy-atom positions.
For example, the MOF with refcode NORYUS contains no hydrogen atoms in the CSD MOF subset and was therefore classified as a hydrogen-error case by LitMOF (see \Cref{fig:noryus}).
Although LitMOF successfully added hydrogen atoms and generated a corrected CIF that is valid in terms of graph representation, the packed arrangement of 4-hydroxypyridine leads to hydrogen atoms being placed too close to each other.
Because the atomic positions of heavy atoms are fixed by experiment, such cases are considered out of scope for correction when constructing a computation-ready experimental structure database.
The same heavy-atom constraint also limits disorder correction.
The three invalid disorder cases found in the manual validation are likewise valid as graph representations but adopt implausible geometries.
In each case the fixed heavy-atom positions cannot assemble a chemically reasonable molecular structure, so the correction is completed with a sound graph but an unsound geometry.
Correcting them properly would require remodeling the structure from scratch or performing a geometry optimization, which would introduce hypothetical structure modeling and make the result a modeled rather than an experimental MOF database, contrary to the construction principle of LitMOF-DB.

From LitMOF-DB (FSR), we further expanded the database by removing or restoring non-framework components (\Cref{fig:litmof_db_result}b).
By removing bound solvent molecules from LitMOF-DB (FSR), we constructed the bound-solvent-removed version, denoted as LitMOF-DB (BSR).
In addition, by restoring charged non-framework components such as counterions, we constructed LitMOF-DB (ION).
\Cref{fig:litmof_db_result}d presents a comparative analysis of these three variants of LitMOF-DB.
As expected, the fraction of porous materials, defined as MOFs with non-zero accessible volume, is highest for BSR, followed by FSR and ION.
The distributions of largest cavity diameter (LCD), pore limiting diameter (PLD), and accessible surface area (ASA) are broadly similar across the three databases, although BSR generally exhibits a right-shifted distribution due to its larger pore sizes.

We compared our curation results with the latest CoRE MOF DB\cite{zhao2025core, zhao_2025_15055758} (\Cref{fig:litmof_db_result}e).
Among the free-solvent-removed CoRE MOF DB entries derived from the CSD, 6,442 structures are labeled as computation-ready (CR) and 8,947 structures as not computation-ready (NCR).
Of the 8,947 NCR structures, LitMOF-DB (FSR) recovered 6,553, corresponding to 73.2~\% of the NCR entries.
The recovery rates for LitMOF-DB (BSR) and LitMOF-DB (ION) are 60.7~\% and 60.6~\%, respectively.
Because the CoRE MOF DB (ASR) contains duplicate entries of the FSR dataset when no bound solvent is detected, such duplicates were removed prior to comparison with LitMOF-DB (BSR).
In addition, LitMOF corrected false-positive computation-ready (CR) structures in the CoRE MOF DB by repairing underlying structural errors, thereby converting them into genuinely computation-ready MOFs.
This correction applied to 15.2~\%, 12.8~\%, and 24.4~\% of the CoRE-labeled CR structures under the BSR, FSR, and ION workflows, respectively.

From the output of the Paper Reader during database construction, we identified 8,771 missing MOFs.
These correspond to MOFs that were synthesized and characterized in the original publications but for which no CIF files were deposited in the CSD.
For each missing MOF, the Paper Reader determines a corresponding parent MOF, defined as the most similar reported structure in the publication, and extracts the transformation relating the parent structure to the missing one.
When available, explanations provided in the publication for the absence of a deposited CIF are also recorded.
The most common transformation types include metal substitution, linker exchange, solvent exchange, and linker functionalization (\Cref{fig:litmof_db_result}c).
The missing-MOF database is maintained as a separate resource from the experimental LitMOF-DB, and its purpose is to recover this otherwise lost information rather than to provide experimental structures.
We retain a missing MOF only when its parent resolves to an accessible CSD refcode with a correct structure.
A CIF is generated only when the transformation from the parent is unambiguous, for example metal substitution at an identified crystallographic site, and these constructed structures are models rather than experimentally determined ones.
The recovered information expands the accessible search space, since these materials were not previously known from structural databases.

\subsection{Structural Errors Distort Direct Air Capture Screening}

\begin{figure}[!ht]
	\centering
	\includegraphics[width=\textwidth]{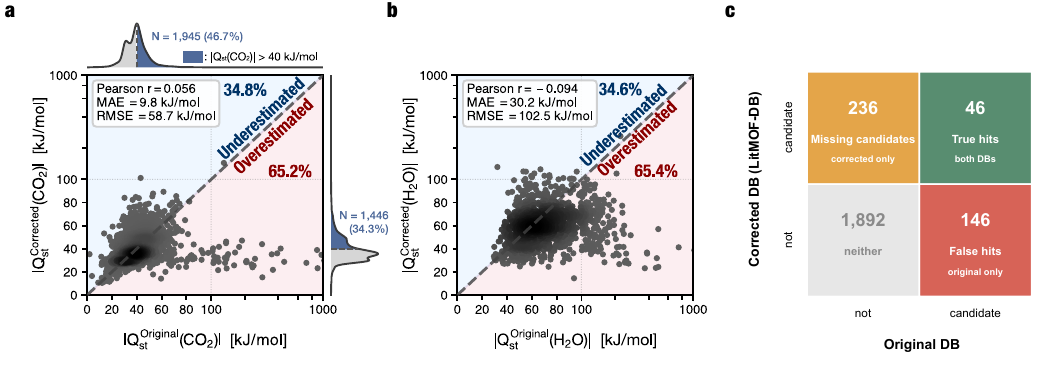}
    \caption{
    \textbf{a}, Comparison of the heats of adsorption of \ce{CO2} calculated from original MOF structures and the corresponding corrected structures in LitMOF-DB.
    \textbf{b}, Same comparison for \ce{H2O}.
    In both panels, deviations from the parity line indicate systematic errors introduced by using uncorrected structures, leading to underestimation or overestimation of adsorption strengths.
    \textbf{c}, Comparison of DAC candidates identified from the original and corrected (LitMOF-DB) structures.
    A structure is counted as a candidate when $|Q_\mathrm{st}(\ce{CO2})| > 40~\mathrm{kJ~mol^{-1}}$ and $S(\ce{CO2}/\ce{H2O}) > 1$ in that database.
    True hits are candidates in both databases, missing candidates only in the corrected database, false hits only in the original database, and the remaining structures fall in neither.
    }
	\label{fig:dac_comparison}
\end{figure}

To further illustrate the impact of structural error, we performed high-throughput screening for direct air capture (DAC) using LitMOF-DB (FSR).
We first identified MOFs that underwent any form of structural correction in LitMOF-DB (FSR), which are labeled as \textit{corrected} in \Cref{fig:dac_comparison}.
For comparison, we also prepared their corresponding original CIFs from the CSD MOF subset in free-solvent-removed form (FSR), labeled as \textit{original} in \Cref{fig:dac_comparison}.
This is the same state on which existing computation-ready databases such as CoRE MOF DB and MOSAEC-DB operate, so the comparison isolates the effect of structural correction rather than that of solvent removal.
Structures unsuitable for DAC were excluded during preprocessing.
Specifically, MOFs with pore-limiting diameters (PLD) smaller than the kinetic diameter of \ce{CO2} (3.3~\AA) were removed.
Structures containing more than 200 atoms per unit cell were also excluded to limit computational cost.
After filtering, 4,256 MOFs were retained for subsequent analysis.
Effective DAC materials require strong \ce{CO2} adsorption at low partial pressures ($\sim$400~ppm) while preferentially adsorbing \ce{CO2} over \ce{H2O}.
To identify candidates satisfying these criteria, we computed the heat of adsorption ($Q_\mathrm{st}$) of \ce{CO2} for both original and corrected MOFs using MLIP (MACE-mh-1 model is used\cite{batatia2025cross}).

As shown in \Cref{fig:dac_comparison}a, the original MOFs exhibit a pronounced tendency to overestimate $|Q_\mathrm{st}|$.
This overestimation includes unphysical values exceeding 100~kJ~mol$^{-1}$, and 97 MOFs yield infinite $|Q_\mathrm{st}|$ values.
These infinite values originate from exceptionally strong computed interaction energies caused by severe structural artifacts in the original MOF structures, which lead to numerical instabilities during adsorption energy evaluation.
In contrast, the distribution of corrected $Q_\mathrm{st}$ values lies within a physically reasonable range, with nearly all values falling between 0 and 100~kJ~mol$^{-1}$.
As a consequence of this overestimation, the number of promising DAC candidates identified using a criterion of $|Q_\mathrm{st}| > 40~kJ~mol^{-1}$ is substantially larger for the original structures (N = 1,945 MOFs) than for the corrected structures (N = 1,446 MOFs).
The Pearson correlation coefficient between the $Q_\mathrm{st}$ values of the original and corrected structures is only 0.056.
This low correlation indicates substantial disagreement in the ranking of candidate materials, rather than merely a difference in the total number of identified candidates.
A similar trend is observed for \ce{H2O}, but with larger errors, as reflected by a mean absolute error (MAE) of 30.2~kJ~mol$^{-1}$ and a root mean square error (RMSE) of 102.5~kJ~mol$^{-1}$ (see \Cref{fig:dac_comparison}b).
These results demonstrate that structural errors in uncorrected MOF databases can severely distort adsorption thermodynamics.

For MOFs satisfying the adsorption-strength criterion of $|Q_\mathrm{st}| > 40~\mathrm{kJ~mol^{-1}}$, we evaluated the dilute-limit \ce{CO2}/\ce{H2O} selectivity,
$S(\ce{CO2}/\ce{H2O})$, defined as the ratio of the Henry's law coefficients of \ce{CO2} and \ce{H2O}.
Structures with $S(\ce{CO2}/\ce{H2O}) > 1$ were therefore considered promising candidates for DAC.
Among the original MOFs, 192 out of 1,945 structures (9.9\%) satisfy this selectivity criterion (\Cref{fig:selectivity_original}).
In contrast, 282 out of 1,446 corrected MOFs (19.5\%) meet the same criterion, despite the smaller number of structures passing the initial $|Q_\mathrm{st}| > 40~\mathrm{kJ~mol^{-1}}$ filter (\Cref{fig:selectivity_corrected}).
This nearly twofold increase indicates that structural correction systematically alters the predicted \ce{CO2}/\ce{H2O} selectivity landscape.

A direct comparison of the DAC candidates from the original and corrected structures is shown in \Cref{fig:dac_comparison}c, with the underlying per-database selectivity distributions provided in \Cref{fig:selectivity_original} and \Cref{fig:selectivity_corrected}.
Only 46 MOFs are identified as candidates in both databases, whereas screening based on the uncorrected structures misses 236 candidates that the corrected database recovers and introduces 146 candidates that the corrected structures do not support.
These discrepancies further highlight how structural errors propagate into selectivity predictions and distort DAC screening outcomes.
Structural correction therefore not only reduces false positives arising from overestimated adsorption strengths but also reveals a substantially larger fraction of MOFs with intrinsically favorable \ce{CO2}/\ce{H2O} selectivity.
Moreover, these recovered candidates are largely unique to LitMOF-DB.
Of the 236 candidates identified only after correction, 175 (74\%) are absent from CoRE MOF DB and 156 (66\%) are absent from MOSAEC-DB (\Cref{tab:dac_provenance}), so a screen based only on existing computation-ready databases would not reach them.
We further confirmed that these conclusions are independent of the energy method, by repeating the screening with a classical force field and benchmarking the dilute-limit Widom insertion against grand-canonical Monte Carlo simulations (\Cref{fig:widom_ff_co2,fig:dac_contingency_ff,fig:gcmc_qst,fig:gcmc_uptake}).

Taken together, these results demonstrate that the use of uncorrected MOF structures can propagate structural errors throughout high-throughput screening workflows, ultimately leading to severely distorted performance metrics and misleading identification of top candidates.
To mitigate this issue, several recently released or updated MOF databases adopt classification-based strategies that attempt to identify and exclude structurally invalid entries prior to screening \cite{zhao2025mofclassifier, gibaldi2025generalizable}.
While such filtering approaches can suppress the most severe artifacts, their effectiveness is fundamentally limited by the accuracy of the classification algorithms, allowing residual errors to persist among the highest-ranked candidates.
More importantly, our results reveal a critical limitation of discard-based strategies.
Many structures labeled as invalid are not intrinsically unsuitable materials, but rather partially corrupted representations that can be recovered through systematic structural correction.
By repairing these structures instead of removing them, LitMOF-DB restores a substantial number of viable candidates that would otherwise be excluded, thereby expanding the accessible materials design space and enabling more reliable, application-driven discovery of MOFs.

\section{\label{analysis}Discussion}
\subsection{Bond Corrected CIF}
Previous computation-ready MOF databases typically discard bond information, so most CIF files in these databases do not contain a usable bond network.
Simulation workflows usually do not require explicit bond definitions, but many structure processing tasks depend critically on them.
Examples include metal node decomposition, organic linker decomposition, and defect generation.
Distance based bond assignment algorithms are commonly used to reconstruct the bond network, yet these methods frequently overestimate or underestimate bonds because MOF bonding environments and periodic structures are highly diverse.
As a result, the reconstructed bond networks are often inaccurate and lead to failures in later structural analysis and modification.

LitMOF-DB resolves this issue by providing CIF files with validated and corrected bond information.
As shown in \Cref{fig:error_correction}a, we apply a bond correction procedure that identifies and fixes wrong bonds while taking periodic boundary conditions and unit cell vectors into account. The CIF files in LitMOF-DB therefore contain a consistent bond network suitable for structure processing and advanced MOF analysis.

\subsection{\label{sec:multi_sol}Multiple Solutions for Correction}
Certain MOF structures admit multiple valid solutions during the correction process.
Such non-unique outcomes occur most often in hydrogen correction and disorder corrections and arise from the inherent limitations of unit cell representations rather than from the LitMOF workflow itself.
Hydrogen atoms are too light to be located reliably through X-ray diffraction or similar crystallographic techniques, so their positions are typically assigned during post-processing.
Our hydrogen-correction procedure follows the same principle.
However, post-processed hydrogen placement does not always have a unique solution.
For example, when a linker contains a carboxylic acid group that is not coordinated to a metal center and the CIF lacks hydrogen atoms on the oxygen sites, a hydrogen atom must be placed on one of the two oxygens.
Either placement is chemically reasonable, and in a MOF with several such groups, the number of valid configurations increases rapidly.
Since crystallographic data cannot distinguish between these possibilities, we select the configuration with the lowest predicted energy.
To manage the potentially large configuration space, we use a foundational MLIP trained on materials to evaluate the candidate structures efficiently.
This approach identifies the most stable hydrogen assignment while keeping the computational cost manageable.
For well-established structural motifs, such as the \ce{\mu3-OH} and \ce{\mu3-O} groups in \ce{Zr6} metal node, predefined chemical rules are applied to reduce ambiguity and further improve the accuracy of the correction.

Disorder correction follows a similar rationale.
Properly resolved disorder in a CIF does not pose any difficulty, but unresolved disorder results in duplicated atoms with fractional occupancies, which prevents the structure from being used reliably in computational simulations.
This situation arises when a portion of the structure has intrinsic degree of freedom in rotation or has specific orientation, so the experimentally supported atomic position cannot be represented by a single configuration in the CIF.
Such dynamic disorder commonly occurs in rotatable organic linkers, functionalized linkers, and weakly bound solvent molecules.
Our disorder correction procedure decomposes these fractional and spatially ambiguous sites into a set of distinct and chemically meaningful configurations.
As illustrated in \Cref{fig:error_correction}c, the meaningful selected configurations can be multiple.
Static disorder, where an atom is replaced by a specific element during correction, can also produce multiple plausible outcomes.
During database construction, we selected the lowest energy configuration using the same strategy applied in hydrogen correction.
However, if requested, the agent can return all possible configurations rather than a single representative structure.

\section{Conclusion}

In this work, we introduced LitMOF, an LLM-driven multi-agent framework that automatically retrieves, interprets, and reconciles information from the literature, crystallographic databases, and CIF files to repair structural errors in MOFs.
Applied to the CSD MOF subset, LitMOF produced LitMOF-DB, a curated collection of 189,567 computation-ready experimental MOFs that resolves long-standing issues of hydrogen placement, bond inconsistencies, and unresolved disorder.
The framework successfully corrected the majority of non-computation-ready CoRE MOF DB structures and also identified 8,771 synthesized but undeposited MOFs by extracting structural information directly from the literature.

Beyond database construction, we demonstrated through a screening targeting direct air capture that structural errors propagate into application-level predictions, systematically distorting adsorption thermodynamics and materials ranking.
These results reveal a fundamental limitation of classification-based workflows that discard invalid structures, as many excluded entries are not intrinsically unsuitable materials but partially corrupted representations that can be recovered through systematic repair.

Together, these findings establish structural repair as a necessary step for reliable data-driven materials discovery.
More broadly, LitMOF provides a generalizable approach for building dynamic, self-correcting materials databases by integrating structured repositories with unstructured scientific text.
As large language models continue to advance, such agentic systems are poised to become core scientific infrastructure for automated materials curation, simulation, and discovery across diverse materials classes.

\section{Methods}

\subsection{LLM Models}
All agents and tools in LitMOF use the Mistral Small 3.2 Instruct (24B) model as the core large language model \cite{mistral_small_3_2_24b}.
PDF documents are processed using the DeepSeek-OCR vision--language model to perform optical character recognition and text extraction \cite{deepseek_ocr_2025}.
The extracted content is converted into a structured markdown format, which is used as input to the LLM.
For other document formats, including HTML and XML, we use a modified version of the parser developed by Kang \textit{et al.} \cite{kang2025harnessing}.

\subsection{MOF Geometric Analysis}
Geometric characterization of the MOF frameworks was carried out with the Zeo++ code \cite{willems2012algorithms}.
For each structure, the pore-limiting diameter (PLD), largest cavity diameter (LCD), accessible surface area, and accessible volume were evaluated.
All quantities were determined by probing the framework with a spherical probe of diameter $1.3~\text{\AA}$, corresponding to the kinetic diameter of \ce{He}.

\subsection{Heat of Adsorption and Selectivity}

The Henry constant ($K_\mathrm{H}$), heat of adsorption ($Q_\mathrm{st}$), and adsorption selectivity were evaluated in the zero-loading limit using the Widom insertion method\cite{widom1963some}.
All ensemble averages were computed at a temperature of $300~\mathrm{K}$.
For each framework, $10{,}000$ Widom insertion trials were performed to ensure statistical convergence.

For each Widom insertion trial, the interaction energy $\Delta U$ was defined as
\begin{equation}
    \Delta U
    =
    U(\mathrm{MOF}+\mathrm{gas})
    -
    U(\mathrm{MOF})
    -
    U(\mathrm{gas}),
\end{equation}
where both the MOF framework and the gas molecule were treated as rigid.
No framework relaxation or guest flexibility was allowed during the insertion procedure.

The heat of adsorption was calculated from the Widom insertion averages as
\begin{equation}
    Q_{\mathrm{st}}
    =
    k_\mathrm{B} T
    -
    \frac{
        \left\langle
        \Delta U \exp\!\left(-\frac{\Delta U}{k_\mathrm{B} T}\right)
        \right\rangle
    }{
        \left\langle
        \exp\!\left(-\frac{\Delta U}{k_\mathrm{B} T}\right)
        \right\rangle
    },
\end{equation}
while the Henry constant was obtained as
\begin{equation}
    K_{\mathrm{H}}
    =
    \frac{
        \left\langle
        \exp\!\left(-\frac{\Delta U}{k_\mathrm{B} T}\right)
        \right\rangle
    }{
        k_\mathrm{B} T
    }.
\end{equation}

Adsorption selectivity between gas species $A$ and $B$ was defined as the ratio of their Henry constants,
\begin{equation}
    S_{A/B}
    =
    \frac{
        K_{\mathrm{H}}^{A}
    }{
        K_{\mathrm{H}}^{B}
    }.
\end{equation}

The interaction energy $\Delta~U$ was evaluated using a pre-trained machine-learning interatomic potential.
Specifically, the MACE-mh-1 model with OMAT PBE output head was employed\cite{batatia2025cross, batatia2022mace}, augmented with a D3 dispersion correction using the Becke--Johnson damping scheme\cite{grimme2010consistent, grimme2011effect, takamoto2022towards}.

\section*{Acknowledgment}
This work acknowledges funding from the National Research Foundation of Korea under project number of RS-2024-00435493.

\newpage
\printbibliography

\end{refsection}

\newpage
\begin{refsection}

\begingroup
\singlespacing
\begin{center}
Supporting Information for

\bigskip
LitMOF: An LLM Multi--Agent for\\
Literature--Validated Metal--Organic Frameworks\\
Database Correction and Expansion

\bigskip
Honghui Kim\textsuperscript{1},
Dohoon Kim\textsuperscript{1},
Jihan Kim\textsuperscript{1*}

\medskip
\textsuperscript{1}Department of Chemical and Biomolecular Engineering,\\
Korea Advanced Institute of Science and Technology, Daejeon, Republic of Korea
\end{center}
\endgroup

\newpage

\begingroup
\singlespacing
\tableofcontents
\endgroup

\newpage

\vspace{2em}

\UseRawInputEncoding

\renewcommand{\thefigure}{S\arabic{figure}}
\renewcommand{\thetable}{S\arabic{table}}
\renewcommand{\theequation}{S\arabic{equation}}
\renewcommand{\thesection}{S\arabic{section}}
\renewcommand{\theHfigure}{SI.\arabic{figure}}
\renewcommand{\theHtable}{SI.\arabic{table}}
\renewcommand{\theHequation}{SI.\arabic{equation}}
\renewcommand{\theHsection}{SI.\arabic{section}}
\setcounter{figure}{0}
\setcounter{table}{0} 
\setcounter{equation}{0}
\setcounter{section}{0}

\vspace{4em}

\newpage

\section{Details of Database Reader}

\begin{figure}[!ht]
	\centering
	\includegraphics[width=\textwidth]{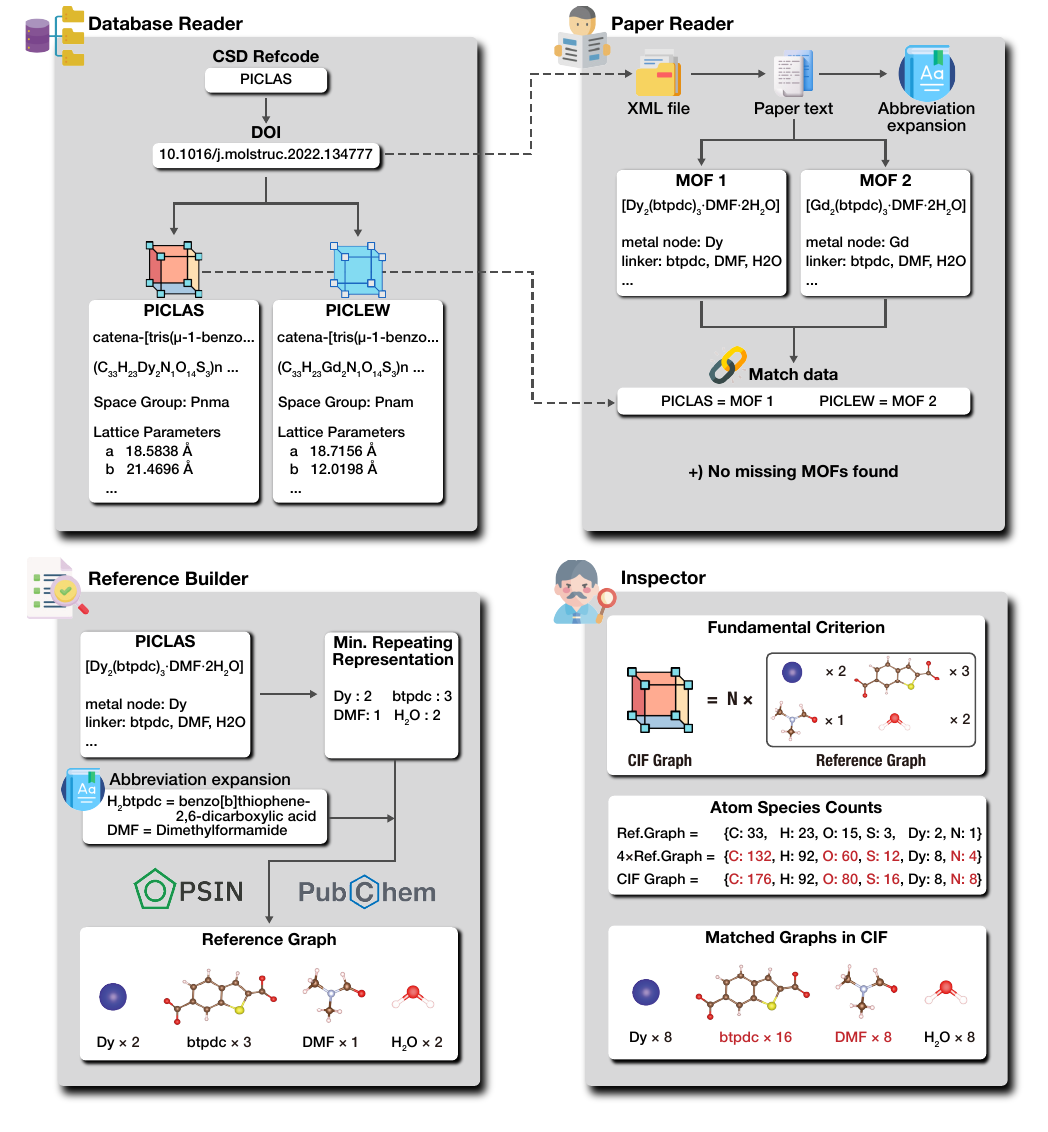}
	\caption{Detailed illustration of how LitMOF retrieves data for the example MOF PICLAS using the Database Reader and Paper Reader, constructs the reference graph using the Reference Builder, and identifies structural errors during inspection of the Inspector \& Editor.}
	\label{fig:piclas_detail}
\end{figure}

\Cref{fig:piclas_detail} gives a detailed worked example of the full LitMOF pipeline for the MOF PICLAS, tracing the data retrieved by the Database Reader and Paper Reader, the reference graph constructed by the Reference Builder, and the structural errors identified by the Inspector \& Editor. The following sections describe each agent in detail.

\subsection{Cambridge Structure Database (CSD)}

One of the main features of the Database Reader is its ability to retrieve available data from the CSD.
Using the CSD Python API, the list of accessible data fields is summarized in \Cref{tab:csd_fields}.

\begin{table}[!ht]
    \centering
    \caption{List of metadata fields retrieved from the Cambridge Structural Database (CSD) by the Database Reader agent.}
    \label{tab:csd_fields}
    \small
    \begin{tabularx}{\textwidth}{@{}>{\ttfamily}l X@{}}
        \toprule
        \normalfont\textbf{Field} & \textbf{Description} \\
        \midrule
        chemical\_name & Chemical name of the MOF structure. \\
        formula & Chemical formula. \\
        synonyms & Alternative names or identifiers. \\
        crystal\_information.crystal\_system & Crystal system (e.g., monoclinic, cubic). \\
        crystal\_information.space\_group & Space group symbol. \\
        crystal\_information.cell\_length\_a,b,c & Unit cell edge lengths ($a$, $b$, $c$). \\
        crystal\_information.cell\_angle\_alpha,beta,gamma & Unit cell angles ($\alpha$, $\beta$, $\gamma$). \\
        crystal\_information.volume & Unit cell volume (\AA$^3$). \\
        has\_disorder & Whether the structure contains disorder. \\
        disorder\_details & Detailed description of disorder, if available. \\
        remarks & Additional remarks or comments. \\
        \bottomrule
    \end{tabularx}
\end{table}

\newpage

\section{Details of Paper Reader}
\subsection{Paper Parsing}
To extract textual information from publications, we employ two complementary parsing methods: (i) direct parsing of XML and HTML sources, and (ii) optical character recognition for PDF documents using a vision--language model.
For XML and HTML files, we build upon the parser originally developed by Kang \textit{et al.}\cite{kang2025harnessing}, with a key modification that preserves inline formatting tags (e.g., boldface), making it easier for the LLM to distinguish MOF identifiers from surrounding text.
Retaining these tags enables the language model to more reliably identify MOF names and other emphasized identifiers in the original documents.
PDF files are processed using DeepSeek-OCR, which provides high accuracy in converting scientific PDFs into structured markdown.
Consequently, XML/HTML inputs result in plain text, whereas PDF inputs yield markdown-style text reflecting the document structure.
All parsed outputs, whether plain text or markdown, are subsequently passed to the language model for downstream extraction and analysis.
Below, \Cref{tab:file_ext_support} summarizes the file formats provided by selected publishers and the corresponding parsing methods used in our workflow.

\begin{table}[!ht]
    \centering
    \caption{List of supported file formats and publishers.}
    \label{tab:file_ext_support}
    \begin{tabular}{lcc}
        \toprule
        \textbf{Publisher} & \textbf{Supported File Format}  & \textbf{Process Method} \\
        \midrule
        Elsevier & XML & Parser \\
        Royal Society of Chemistry & HTML & Parser \\
        American Chemical Society & XML & Parser \\
        Wiley & PDF & Vision Language Model \\
        International Union of Crystallography & PDF & Vision Language Model \\
        \bottomrule
    \end{tabular}
\end{table}

Of the 53,306 publications linked to the CSD MOF subset entries, we accessed
36,146. The remaining publications are from publishers that do not provide full
text for text mining.

\newpage
\subsection{Full-Document LLM Inference vs.\ Retrieval-Augmented Generation (RAG)}
\label{sec:rag_vs_full}
\begin{table}[!ht]
\centering
\small
\caption{Direct comparison of raw information extracted for MOF1 of DOI 10.1021/ic100164k using full-document inference and retrieval-augmented generation (RAG).}
\label{tab:rag_comparison}
\setlength{\tabcolsep}{5pt}
\renewcommand{\arraystretch}{1.15}
    \begin{tabularx}{\textwidth}{lXX}
    \toprule
    Field & \textbf{Full document inference} & \textbf{RAG} \\
    \midrule
    identifier\_in\_text & 1 & 1 \\
    structural\_formula & Co$_2$(4,4'-bpy)(tfhba)$_2\cdot$4,4'-bpy & -- \\
    chemical\_formula & CoC$_{17}$H$_8$F$_4$N$_2$O$_3$ & -- \\
    crystal\_system & orthorhombic & -- \\
    space\_group & Pmna & -- \\
    cell\_parameters & $a,b,c=26.271,8.9909,6.4018$~\AA; $\alpha=\beta=\gamma=90^\circ$ & -- \\
    volume & 1512.1~\AA$^3$ & -- \\
    metal\_node & Co & Co \\
    metal\_oxidation\_state & 2+ & -- \\
    organic\_linker & 4,4'-bpy, tfhba & 4,4'-bpy, tfhba \\
    solvent & -- & -- \\
    important\_notes & Contains one-dimensional chains of corner-sharing CoO$_4$N trigonal bipyramids. Uncoordinated 4,4'-bpy molecule fills the cavity. & Contains a one-dimensional chain of corner-sharing MO$_4$N trigonal bipyramids. Isostructural with MOF2. \\
    \bottomrule
    \end{tabularx}
\end{table}

The Paper Reader is designed to extract structured information from scientific papers and answer queries based on their content.
Because a full research article can contain a nontrivial amount of text, providing the entire document as input to an API-based LLM is sometimes considered inefficient in terms of cost or context usage.
A common strategy to address this concern is Retrieval-Augmented Generation (RAG), which partitions a document into smaller text segments, retrieves potentially relevant portions, and generates answers based only on the retrieved context.

In our application, however, RAG is not well suited for several reasons.
First, the full text of a typical chemistry paper is on the order of 10,000 tokens, which comfortably fits within the context window of modern LLMs that support 100,000 tokens or more.
Second, the information required for reliable materials data extraction is often distributed across the manuscript, including the main text, tables, figure captions, and crystallographic descriptions.
When a document is partitioned into independent chunks, structural cues, cross-references, and globally defined identifiers are frequently separated, which can degrade extraction accuracy even if the retrieved chunks are locally relevant.

To quantitatively compare full-document inference with RAG, we conducted a controlled experiment using the same language model and identical prompts for both approaches, differing only in the inference strategy.
The results are summarized in \Cref{tab:rag_comparison}.
As a representative case, we analyzed the paper with DOI 10.1021/ic100164k, which reports the synthesis of two distinct MOFs; the table presents the extraction results for the first structure (MOF1).
For the RAG-based approach, the document was divided into overlapping segments of 512 tokens with a 20\% overlap, yielding 43 chunks in total.
For each query, a similarity-based retriever selected the six most relevant chunks using vector embeddings generated with the \texttt{nomic-embed-text} model\cite{nussbaum2024nomic}, and the selected chunks were provided to the LLM as contextual input.

Despite the use of identical prompts and comparable textual content, the RAG-based approach failed to recover several key structural and crystallographic descriptors that were successfully extracted via full-document inference.
In particular, complete chemical formulas, crystallographic parameters, and detailed structural descriptions were frequently missing or only partially specified in the RAG outputs.
This behavior reflects the difficulty of reconstructing a coherent, document-level representation of a material when relevant information is fragmented across multiple retrieved segments.

Taken together, these results indicate that, for single-paper information extraction tasks in which the full document fits within the model context window, full-document inference yields more complete and chemically consistent results than RAG.
While RAG remains effective for large-scale or multi-document retrieval problems, it introduces unnecessary complexity and accuracy loss for the document-level reasoning required in this work.
Accordingly, we adopt the Mistral Small 3.2 Instruct 24B model, an open-source LLM with a 128K-token context window, to enable direct full-document inference throughout this study.

\newpage

\subsection{Dynamic Prompt}
\label{sec:dynamic_prompt}
To improve and validate the output of data extraction in the Paper Reader, we adopt dynamic prompting.
Dynamic prompting adjusts the instruction and context provided to the LLM based on intermediate results, enabling the model to focus on unresolved or ambiguous information.
This adaptive formulation enhances extraction accuracy, especially in cases where the initial response is incomplete or where multiple candidate interpretations exist in the source text.

The core tasks of the Paper Reader are threefold: 
(1) \textbf{[Extraction]} extracting synthesized MOF data from the text, 
(2) \textbf{[Matching]} aligning the reported MOFs with their corresponding CSD reference codes, and 
(3) \textbf{[Detection]} identifying missing MOF entries. 
Dynamic prompting enhances the reliability of each task by iteratively refining queries based on inconsistencies or incomplete information. 
Here, we illustrate how dynamic prompting addresses these three representative tasks.

\begin{algorithm}[H]
    \setstretch{1.1} 
    
    \caption{\textbf{[Extraction]} Extracting MOF information from a paper}
    \DontPrintSemicolon
    
    \KwInput{$T_{\text{paper}}$ (paper text), $I_{\text{csd}}$ (CSD information)}
    \KwOutput{$I_{\text{mof}}$ (set of extracted MOF entries)}

    $M \leftarrow []$ \tcp*[r]{Init. empty messages}
    
    \BlankLine
    $P_{\text{read}} \leftarrow \GenRead(T_{\text{paper}})$\; 
    $M \leftarrow M + [ \UserMsg(P_{\text{read}}) ]$\;
    $I_{\text{mof}} \leftarrow \LLMJson(M)$ \tcp*[r]{invoke LLM with JSON output format}
    
    \BlankLine
    \If{$|I_{\text{mof}}| < |I_{\text{csd}}|$ \tcp*[r]{Retry if counts mismatch}}{ 
    
        $M \leftarrow M + [ \AIMsg(I_{\text{mof}}) ]$\;
        
        $P_{\text{retry}} \leftarrow \GenRetry(|I_{\text{mof}}|, I_{\text{csd}})$\;
        $M \leftarrow M + [ \UserMsg(P_{\text{retry}}) ]$\;
        
        $R_{\text{reason}} \leftarrow \LLM(M)$ \tcp*[r]{invoke LLM to reason the mismatch}
        $M \leftarrow M + [ \AIMsg(R_{\text{reason}}) ]$\;
        
        $P_{\text{struct}} \leftarrow \text{StructuredPromptOnly}$\;
        $M \leftarrow M + [ \UserMsg(P_{\text{struct}}) ]$\;
        
        $I_{\text{mof}} \leftarrow \LLMJson(M)$ \tcp*[r]{update the $I_{\text{mof}}$}
    }
    
    \KwRet{$I_{\text{mof}}$}\;
\end{algorithm}

\begin{promptbox}[title=$P_{read}$ \textbf{(paper\_read\_prompt)}]
\begin{lstlisting}[language=prompt]
You are an expert assistant in reading scientific publications and extracting structured information about synthesized metal-organic frameworks (MOFs) or other crystalline materials.

Your task:
1. Read the provided text from a scientific paper.
2. Identify each synthesized MOF or crystal in the order they appear. Do NOT include MOFs that are only mentioned in the Introduction and not synthesized in this paper.
3. Produce a JSON object whose top-level keys are "MOF1", "MOF2", "MOF3", ... corresponding to each distinct synthesized MOF.
4. For each MOF, extract the following fields when present in the text (leave empty string "" otherwise):
   - identifier_in_text: the label the paper uses (e.g., "1", "2-Ln", "1-Eu").
   - structural_formula: e.g. [{Dy(H3C-Im-CH2COO)4(H2O)}](PF6)3⋅2H2O
   - chemical_formula: e.g. C44H33Cd2N6O15
   - crystal_system: e.g. monoclinic, triclinic, orthorhombic
   - space_group: e.g. P21/c, C2/c
   - cell_parameters: e.g. a = 10.123 Å, b = 12.456 Å, c = 15.789 Å, alpha = 90°, beta = 110.5°, gamma = 90°
   - volume: e.g. 1234.56 Å3
   - metal_node
   - metal_oxidation_state
   - organic_linker (list)
   - solvent (list)
   - important_notes: structural quirks, disorder remarks, relevant details.
5. Add a top-level key "abbreviations" mapping each abbreviation used in the paper to its full chemical name (e.g., {"L": "3-(2-pyridyl)pyrazole"}).
6. If an organic linker abbreviation cannot be parsed to SMILES, add any descriptive hint to that MOF's important_notes.
7. Return only valid JSON. No commentary outside the JSON.

PAPER CONTEXT:
|{paper_text}|
\end{lstlisting}
\end{promptbox}

\begin{promptbox}[title=$P_{retry}$ \textbf{(paper\_read\_retry\_prompt)}]
\begin{lstlisting}[language=prompt]
Why are there fewer MOFs extracted than expected?
You found |{num_mofs}| MOFs, but the CSD connected to this paper has |{expected_num_mofs}| entries - so I expect |{expected_num_mofs}|.

Here is the CSD entry list and metadata:
|{csd_info}|

Check this information carefully and see if any synthesized MOF was missing from your previous extraction. Think step by step.
\end{lstlisting}
\end{promptbox}

\begin{promptbox}[title=$P_{struct}$ \textbf{(paper\_read\_retry\_structured\_prompt)}]
\begin{lstlisting}[language=prompt]
Based on the reasoning above, return an updated PAPER_INFO JSON object that now includes any MOFs you originally missed.

Return only valid JSON.
\end{lstlisting}
\end{promptbox}

\newpage

\begin{algorithm}[H]
    \setstretch{1.1}
    
    \caption{\textbf{[Matching]} Matching extracted MOFs with CSD reference codes}
    \DontPrintSemicolon
    
    \KwInput{$T_{\text{paper}}$ (paper text), $I_{\text{csd}}$ (CSD information), $I_{\text{mof}}$ (extracted MOFs)}
    \KwOutput{$I_{\text{match}}$ (mapping of CSD codes to MOF IDs)}

    $M \leftarrow []$ \tcp*[r]{Init. empty messages}
    
    \BlankLine
    $P_{\text{match}} \leftarrow \GenMatch(I_{\text{mof}}, I_{\text{csd}}, T_{\text{paper}})$\;
    $M \leftarrow M + [ \UserMsg(P_{\text{match}})]$\;
    $I_{\text{match}} \leftarrow \LLMJson(M)$\; 
    
    $P_{\text{refine}} \leftarrow \GenRefine(I_{\text{mof}}, I_{\text{match}})$\;
    $I_{\text{match}} \leftarrow \LLMJson( [ \UserMsg(P_{\text{refine}}) ] )$\;

    $M \leftarrow M + [\AIMsg(I_{\text{match}})]$\;
    
    \BlankLine
    $U_{\text{miss}} \leftarrow \GetUnmatched(I_{\text{match}})$ \tcp*[r]{Identify missing matches (refcodes)}
    
    \If{$U_{\text{miss}} \neq \emptyset$ \tcp*[r]{Retry if matches are missing}}{
    
        $P_{\text{why}} \leftarrow \text{``Why not match: ''} + \text{Join}(U_{\text{miss}})$\;
        $M \leftarrow M + [ \UserMsg(P_{\text{why}}) ]$\;
        
        $R_{\text{reason}} \leftarrow \LLM(M)$ \tcp*[r]{invoke LLM to reason the mismatch}
        $M \leftarrow M + [ \AIMsg(R_{\text{reason}}) ]$\;
        
        $P_{\text{retry\_match}} \leftarrow \text{MatchRetryPrompt}$\;
        $M \leftarrow M + [ \UserMsg(P_{\text{retry}}) ]$\;
        
        $I_{\text{match}} \leftarrow \LLMJson(M)$ \tcp*[r]{update the $I_{\text{match}}$}
    }
    
    \KwRet{$I_{\text{match}}$}\;
\end{algorithm}

\begin{promptbox}[title=$P_{match}$ (\textbf{paper\_match\_prompt})]
\begin{lstlisting}[language=prompt]
You have two JSON inputs:

- PAPER_INFO : summary of MOFs in the paper. Keys are "MOF1", "MOF2", ...; their real labels in the paper are in the "identifier_in_text" field.
- CSD_INFO   : CSD metadata, keys are reference codes (e.g. "CIBYIX", "CIBYOD").

Your task: produce a flat JSON object whose keys are each CSD reference code, and whose values are the matching PAPER_INFO label.

CRITICAL: the output KEYS must be the CSD reference codes EXACTLY as they appear as keys in CSD_INFO (e.g. "CIBYIX", "XUVPIP") - copy them verbatim. NEVER use paper compound numbers ("1", "2") or PAPER_INFO labels ("MOF1") as keys. Keys = refcodes, values = MOF labels.

- If you are certain MOF1 corresponds to CIBYIX, map "CIBYIX": "MOF1".
- If unsure, map the refcode to "" (empty string).

Use every available signal:
- cell parameters
- crystal system
- space group
- metal type
- organic linker
- solvent / counter-ion
- chemical formula
- structural formula
- oxidation state

PAPER_INFO:
|{paper_info}|

CSD_INFO:
|{csd_info}|

Paper text (for reference):
|{paper_context}|

Values MUST be either a PAPER_INFO label like "MOF1" or "". Return only the mapping JSON. Think step by step.
\end{lstlisting}
\end{promptbox}

\begin{promptbox}[title=$P_{refine}$ (\textbf{paper\_match\_refine\_prompt})]
\begin{lstlisting}[language=prompt]
Refine the previous MATCH_INFO. A common mistake is having values like the real "identifier_in_text" instead of the PAPER_INFO label ("MOF1", "MOF2", ...).

PAPER_INFO:
|{paper_info}|

MATCH_INFO:
|{match_info}|

Refined values must be exactly one of: |{paper_info_keys}| - or "" for unsure.

Return only the refined mapping JSON. Think step by step.
\end{lstlisting}
\end{promptbox}

\begin{promptbox}[title=$P_{retry\_match}$ (\textbf{paper\_match\_retry\_prompt})]
\begin{lstlisting}[language=prompt]
Based on your reasoning about why those refcodes were not matched, return an updated MATCH_INFO JSON that includes plausible matches for them.

Return only the mapping JSON.
\end{lstlisting}
\end{promptbox}

\newpage

\begin{algorithm}[H]
    \setstretch{1.1}
    
    \caption{\textbf{[Detection]} Identifying missing MOF entries in CSD database}
    \DontPrintSemicolon
    
    \KwInput{$T_{\text{paper}}$ (paper text), $I_{\text{csd}}$ (CSD info), $I_{\text{mof}}$ (extracted MOFs), $I_{\text{match}}$ (match results)}
    \KwOutput{$I_{\text{miss}}$ (details of missing MOFs)}

    $M \leftarrow []$ \tcp*[r]{Init. empty messages}
    $I_{\text{miss}} \leftarrow \emptyset$\;
    
    \BlankLine
    $S_{\text{matched}} \leftarrow \GetMatched(I_{\text{match}})$ \tcp*[r]{Get matched refcodes}
    
    $S_{\text{diff}} \leftarrow \Keys(I_{\text{mof}}) \setminus S_{\text{matched}}$ \tcp*[r]{Identify unmatched refcodes}
    
    \BlankLine
    \If{$S_{\text{diff}} \neq \emptyset$ \tcp*[r]{Proceed only if missing MOFs exist}}{
        
        $P_{\text{detect}} \leftarrow \GenDetect(I_{\text{mof}}, I_{\text{csd}}, I_{\text{match}}, S_{\text{diff}})$\;
        $M \leftarrow M + [ \UserMsg(P_{\text{detect}}) ]$\;
        
        $I_{\text{miss}} \leftarrow \LLMJson(M)$ \tcp*[r]{invoke LLM to detail missing MOFs}
        
    }    
    \KwRet{$I_{\text{miss}}$}\;
\end{algorithm}

\begin{promptbox}[title=$P_{detect}$ (\textbf{paper\_missing\_mof\_prompt})]
\begin{lstlisting}[language=prompt]
You are identifying missing MOF structures from a scientific publication.

Definition: a "missing MOF" is one that the paper synthesized and characterized but for which no CIF was deposited in the CSD (no independent reference code).

Inputs:
- PAPER_INFO : extractor output (keys "MOF1", "MOF2", ...).
- CSD_INFO   : CSD metadata keyed by reference code.
- MATCH_INFO : for each PAPER_INFO label, either the matched refcode or "".

Tasks:
1. List every missing MOF - present in PAPER_INFO but not matched to any CSD refcode.
2. For each missing MOF, choose a plausible parent_mof by comparing metal, linker, solvent / counter-ion, or topology with:
   - any PAPER_INFO entry NOT marked missing, or
   - a well-known prototype explicitly named in the text.
3. Describe the minimal transformation from parent to missing. Allowed categories (use these exact tokens):
   - metal_substitution
   - linker_exchange
   - linker_functionalisation
   - solvent_modification
   - other
4. If the paper states why no CIF was deposited (e.g. "poor crystallinity", "disordered, not solved", "already reported elsewhere"), quote that as reason_no_cif. Otherwise leave it as "".

PAPER_INFO:
|{paper_info}|

CSD_INFO:
|{csd_info}|

MATCH_INFO:
|{match_info}|

Paper text:
|{paper_context}|

There are |{missing_num}| missing MOF(s). Unmatched: |{missing_mofs}|.

Output: a JSON object with top-level keys = the missing MOF labels (e.g. "MOF3"), and each value an object with:
  "identifier_in_text"     : string
  "parent_mof"            : string
  "transformation"         : one of the 5 tokens above
  "transformation_details" : one-sentence explanation
  "reason_no_cif"          : quoted reason or ""

If there are no missing MOFs, return an empty JSON object. Think step by step.
\end{lstlisting}
\end{promptbox}

\begin{table}[!htbp]
    \centering
    \caption{Metadata fields extracted by the Paper Reader agent.}
    \label{tab:paper_reader_descrip}
    \begin{tabularx}{\textwidth}{
        >{\RaggedRight\arraybackslash}p{3.2cm}
        >{\RaggedRight\arraybackslash}p{6.0cm}
        X
    }
        \toprule
        \textbf{Field group} & \textbf{Key} & \textbf{Description} \\
        \midrule
        \textbf{Paper information} & \texttt{.MOF*.identifier\_in\_text} & Label used for the MOF in the text (e.g., ``1-Ln'', ``1-Eu''). \\
        & \texttt{.MOF*.structural\_formula} & Structural formula as reported in the paper. \\
        & \texttt{.MOF*.chemical\_formula} & Chemical formula of the MOF. \\
        & \texttt{.MOF*.crystal\_system} & Crystal system (e.g., triclinic, monoclinic). \\
        & \texttt{.MOF*.space\_group} & Space group symbol. \\
        & \texttt{.MOF*.cell\_parameters} & Unit-cell parameters ($a$, $b$, $c$, $\alpha$, $\beta$, $\gamma$). \\
        & \texttt{.MOF*.volume} & Unit-cell volume. \\
        & \texttt{.MOF*.metal\_node} & Composition of the metal node (e.g., Eu, Tb, or mixture). \\
        & \texttt{.MOF*.metal\_oxidation\_state} & Oxidation state of the metal node(s). \\
        & \texttt{.MOF*.organic\_linker} & Name and formula of the organic linker. \\
        & \texttt{.MOF*.solvent} & Main solvent(s) used in synthesis or structure description. \\
        & \texttt{.MOF*.important\_notes} & Remarks on structure or properties (e.g., luminescence, applications). \\
        & \texttt{.abbreviations[*]} & Dictionary mapping abbreviations (e.g., H2L, DMA) to full names. \\
        \midrule
        \textbf{Match information} & \texttt{[refcode]} & Mapping between a CSD refcode (e.g., \texttt{GAZDEU}) and its corresponding MOF ID (e.g., \texttt{MOF3}). \\
        \midrule
        \textbf{Missing MOFs} & \texttt{.MOF*.identifier\_in\_text} & Label of a MOF mentioned in the paper but absent from structural databases. \\
        & \texttt{.MOF*.parent\_mof} & Parent MOF entry used as the structural reference. \\
        & \texttt{.MOF*.transformation} & Type of transformation from the parent MOF (e.g., metal substitution). \\
        & \texttt{.MOF*.transformation\_details} & Description of how the missing MOF differs from its parent structure. \\
        & \texttt{.MOF*.reason\_no\_cif} & Reason for the absence of a CIF file, if provided. \\
        \bottomrule
    \end{tabularx}
\end{table}

\newpage

\section{Details of Reference Builder}

\begin{figure}[!ht]
	\centering
	\includegraphics[width=\textwidth]{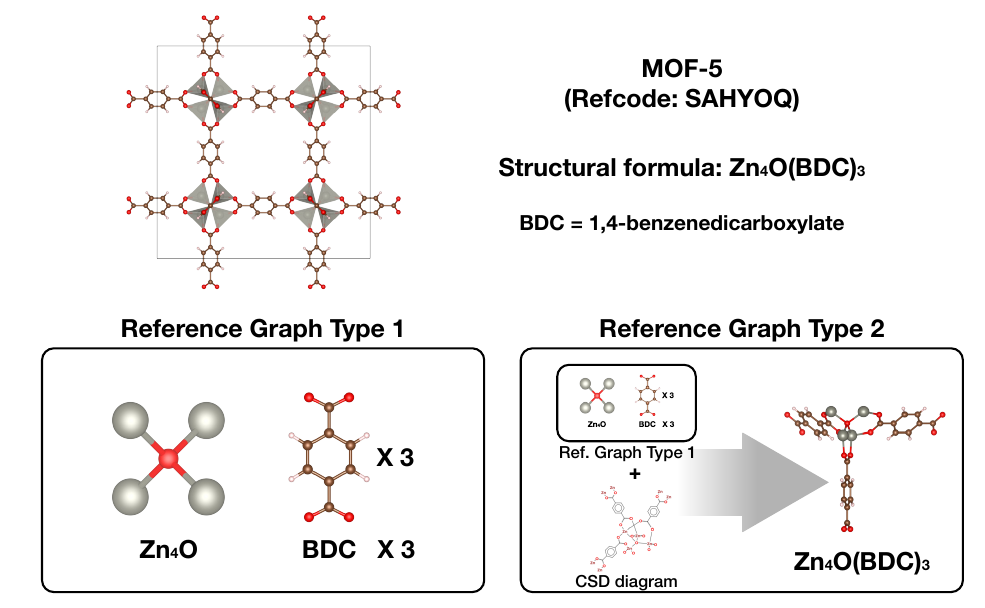}
	\caption{Two types of reference graph.}
	\label{fig:ref_graph}
\end{figure}

\begin{figure}[!ht]
	\centering
	\includegraphics[width=\textwidth]{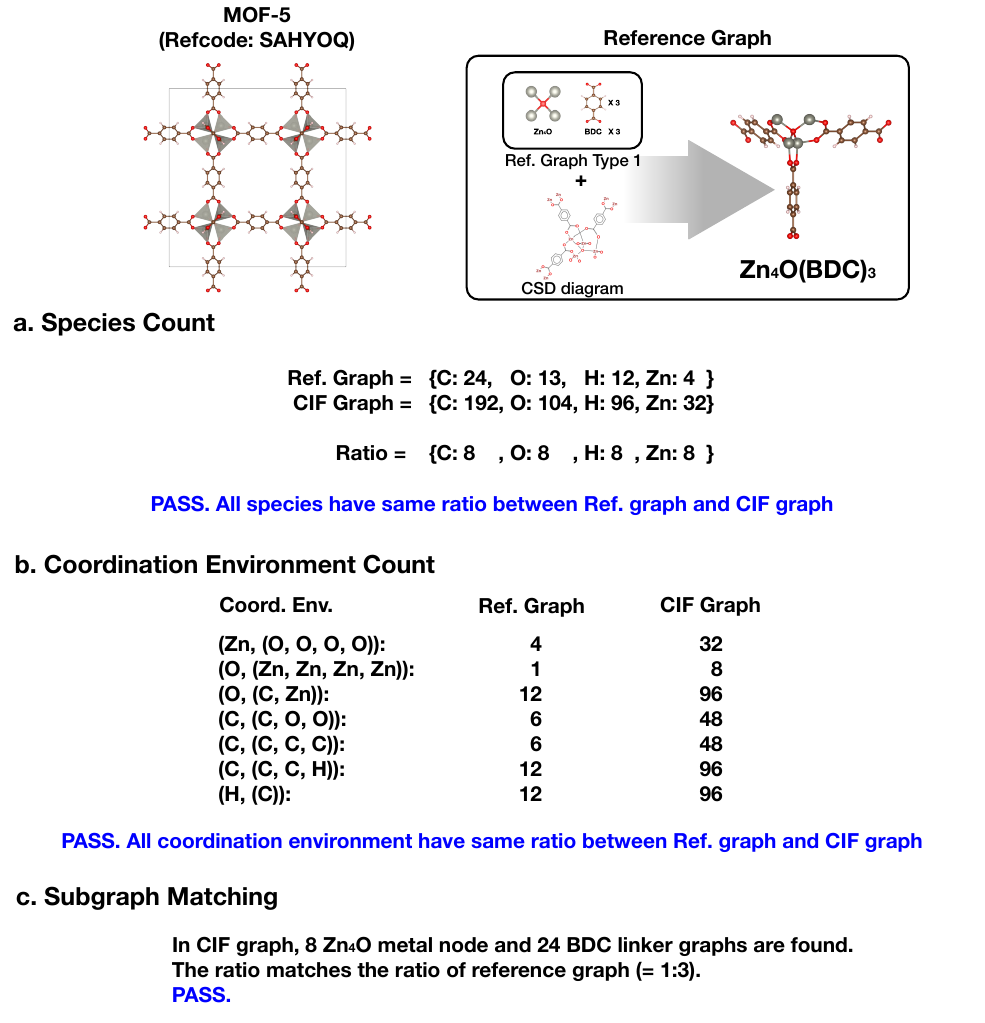}
	\caption{Three types of comparison test between reference graph and CIF graph.}
	\label{fig:comparison_test}
\end{figure}

Building a correct representative structure of a MOF is a key process of structure correction and validation, and we call it as reference graph.
The sources of reference graph are paper and database.
In the case of paper, the Reference Builder utilizes the structural formula and dictionary of abbreviation expansion to get full name of each component of the structural formula.
Then, the text type names are transformed to atomic graph objects by using PubChem API, IUPAC name parser (pyOpsin), and pre-defined name to SMILES mapping.
In the case of database, the Reference Builder utilizes chemical diagram of a MOF provided by CSD.
Using all these information, Reference Builder cross-validates each other and make the final reference graph (see reference graph type 2 in \Cref{fig:ref_graph}).
Once built, the reference graph is compared against the CIF graph at three levels of structural detail to localize any discrepancies (\Cref{fig:comparison_test}).

When the reference graph is built from the paper-side structural formula, the Reference
Builder uses the prompt below to split the structural formula into metal node, organic
linker, and solvent components and to resolve abbreviations against the paper's
abbreviation table.

\begin{promptbox}[title=$P_{cat}$ \textbf{(reference\_categorize\_prompt)}]
\begin{lstlisting}[language=prompt]
You are an expert assistant categorizing MOF components from their structural
formula. For each MOF below, split the structural_formula into three groups
and an integer ratio:

- metal_node      : central metal cation(s), usually inside [...].
- organic_linker  : coordinating ligands (carboxylates, imidazoles, pyridines,
  etc.), also usually inside [...]. Coordinated water (e.g. H2O aqua
  ligands) belongs here.
- solvent         : counterions and lattice solvent, usually written outside
  [...] or after a dot. Common: PF6, ClO4, BF4, NO3, H2O, DMF, MeOH.

Rules:
- Do NOT invent components. Use the names exactly as they appear in the
  formula, except:
  - Strip oxidation-state / charge markers: Fe(III) -> Fe, SO4 2- -> SO4.
  - Resolve linker abbreviations to the full name from the paper's
    abbreviations table when one is given. BDC -> benzene-1,4-dicarboxylate.
  - Keep these short notations as-is: H2O, PF6, BF4, ClO4, DMF, MeOH.
  - Exceptions that ARE expanded: CN -> cyanide, SO4 -> sulfate,
    NO3 -> nitrate, CO3 -> carbonate, PO4 -> phosphate, NH4 -> ammonium.
- A component (e.g. H2O) may appear in BOTH organic_linker AND solvent
  when the formula shows it bound and free. Split the count accordingly.
- Ratios are positive integers.

Output is a single object keyed by CSD refcode. Each value has the three
component maps (which may be empty).

Refcode to structural formula:
|{structural_formula}|

Paper abbreviations:
|{abbreviations}|
\end{lstlisting}
\end{promptbox}

\newpage
\section{Details of Inspector \& Editor}

The Inspector \& Editor applies the structural corrections (bond, hydrogen, disorder, in
that order), validates each step against the reference framework graph, and decides the
next action when validation fails. When a correction finishes but final validation fails
or produces suboptimal output, the Inspector runs a post-mortem analysis with the prompt
below. The analysis reads the failure metadata and returns a single most plausible cause
and a recommended next action, which routes the entry to a rebuild, a retry with a
different method, manual review, partial acceptance, or abort.

\begin{promptbox}[title=$P_{ana}$ \textbf{(inspector\_analyze\_failure\_prompt)}]
\begin{lstlisting}[language=prompt]
You are the LitMOF inspector's post-mortem analyst. A structure correction
pipeline just finished for a single CSD refcode but final validation failed
or produced suboptimal output. Your job is to read the failure metadata and
pick the most plausible cause and the next recommended action.

Context
- The pipeline tries (in order): bond correction -> hydrogen correction ->
  disorder correction. Each step calls a tool-layer algorithm (csg.correct_*)
  and immediately validates against the reference framework graph built from
  the paper's chemical diagram.
- Validation returns three flags:
  - lowlevel_result: pass (heavy atoms + H match), h_diff (heavy atoms
    match but H counts differ), or diff (heavy atoms disagree).
  - detailed_match: whether per-atom coordination environments match.
  - correction_successful: bool.
- A pristine output requires lowlevel_result == pass AND detailed_match == True.

Causes (pick ONE):
- reference_graph_wrong  : the framework graph built by reference_builder is
  itself wrong (e.g. CSD's chemical diagram is inconsistent with the deposited
  atoms, or paper-side nomenclature introduced a structural artifact).
- tool_algorithm_limit   : the tool layer's correction algorithm reached its
  limit on this case (e.g. bond_correction cascade failed even though the
  reference looked sane; correct_hydrogen graph-matching ambiguity).
- paper_nomenclature_error : paper's structural_formula or abbreviations table
  contains errors that propagated into the reference (e.g. phenyl written for
  what should be benzene; H counts inconsistent with the actual ligand).
- csd_data_corruption    : CSD entry itself appears corrupt (missing atoms,
  malformed bonds, encoding issues).
- unknown                : none of the above clearly applies.

Recommended next actions (pick ONE):
- rebuild_reference  : re-run reference_builder, possibly forcing different name
  resolution variants or rejecting the corrupt component.
- retry_other_method : same tool-layer step with a different method argument
  (e.g. switch correct_hydrogen from identity_mapping to graph_matching).
- flag_for_manual    : human review required; data side likely needs patching.
- accept_partial     : current state is the best automated output achievable.
- abort              : refuse to touch this entry; downstream should skip.

Inputs
Refcode: |{refcode}|
Diagnosis (initial):            |{diagnosis}|
Routing taken:                  |{routing}|
Per-step validation results:    |{step_results}|
Final validation:               |{final_validation}|
Suspected H issues hint:        |{suspected_h_issues}|

Output
Return a CorrectionFailureAnalysis with: suspected_cause, explanation (2-4
sentences citing concrete numbers from the metadata above),
recommended_next_action, and confidence (0.0-1.0). Be concrete; quote species
counts, lowlevel_result strings, or other observable evidence when you can.
\end{lstlisting}
\end{promptbox}

We illustrate the corrected output of the Inspector \& Editor with two examples.

\begin{figure}[!htbp]
	\centering
	\includegraphics[width=0.9\textwidth]{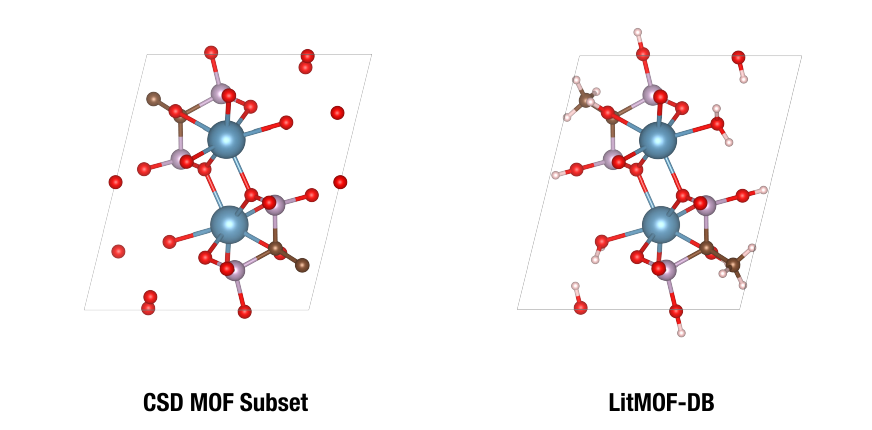}
	\caption{Corrected image of MOF LODHAP.}
\end{figure}

\begin{figure}[!htbp]
	\centering
	\includegraphics[width=0.9\textwidth]{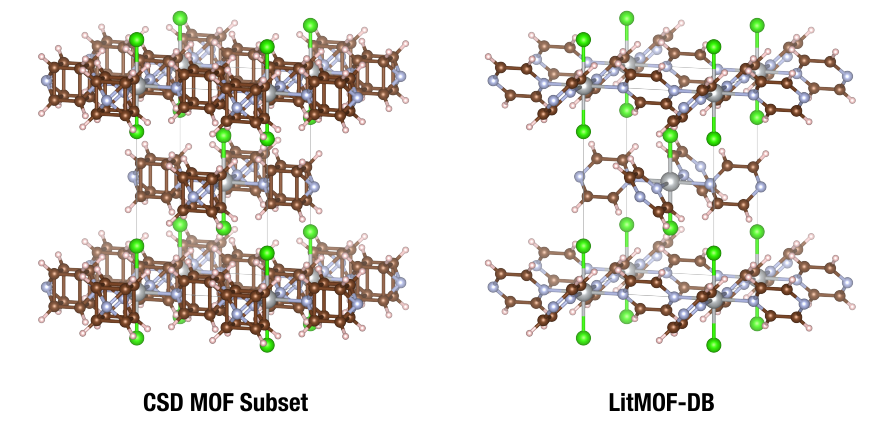}
	\caption{Corrected image of MOF EKOFER.}
\end{figure}

\subsection{MLIP Energy Selection for Disorder Candidates}

\begin{figure}[!htbp]
	\centering
	\includegraphics[width=0.7\textwidth]{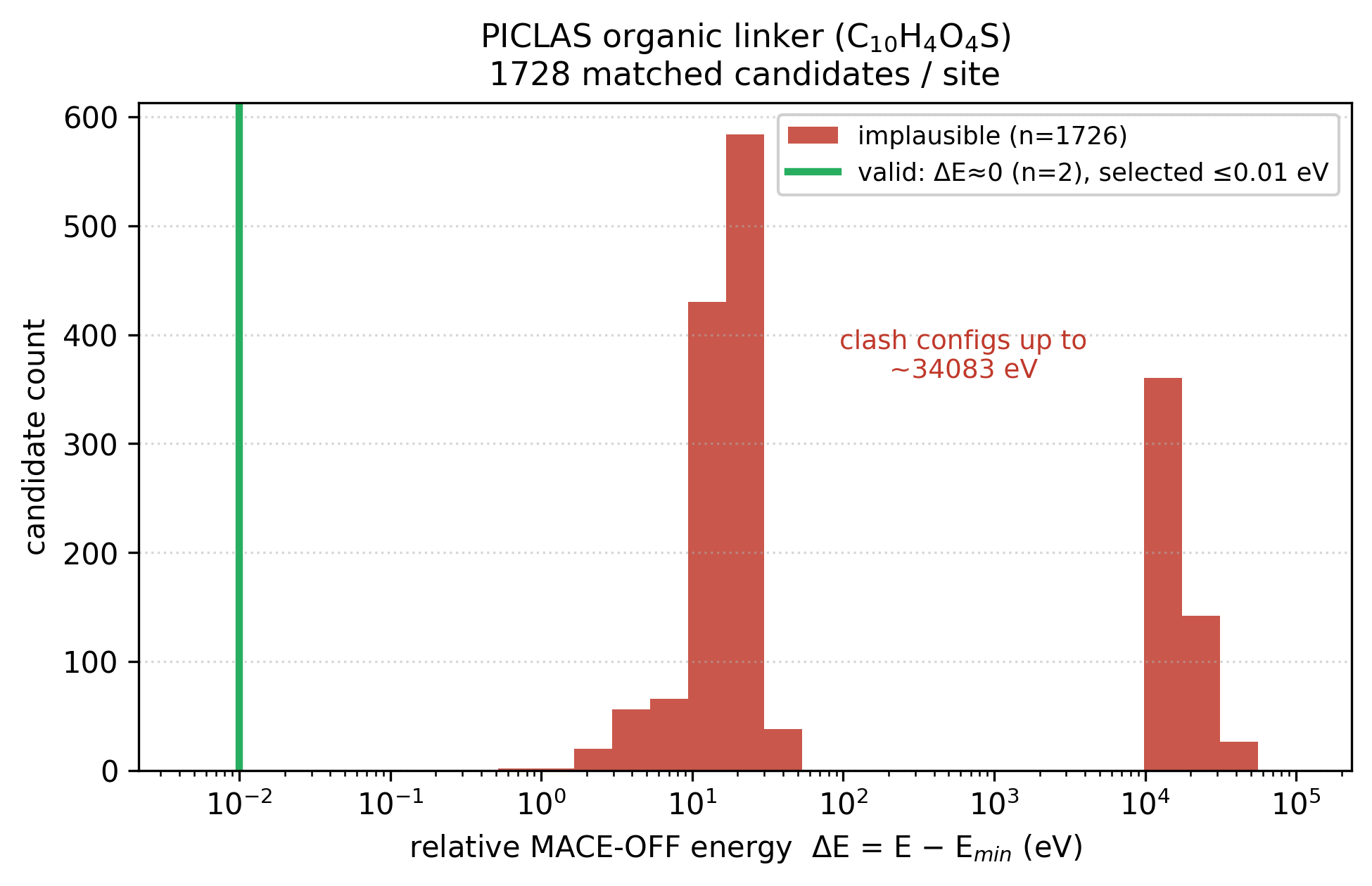}
	\caption{Relative MLIP energy ($\Delta E = E - E_\mathrm{min}$) of the 1,728
	matched candidates for the organic linker (\ce{C10H4O4S}) of PICLAS. The two
	valid configurations ($\Delta E \approx 0$) are clearly separated from the
	1,726 implausible candidates, which extend up to $\sim$34,000~eV for clashing
	configurations.}
	\label{fig:piclas_ehisto}
\end{figure}

\begin{figure}[!htbp]
	\centering
	\begin{minipage}{0.35\textwidth}
		\centering
		\includegraphics[width=0.8\linewidth]{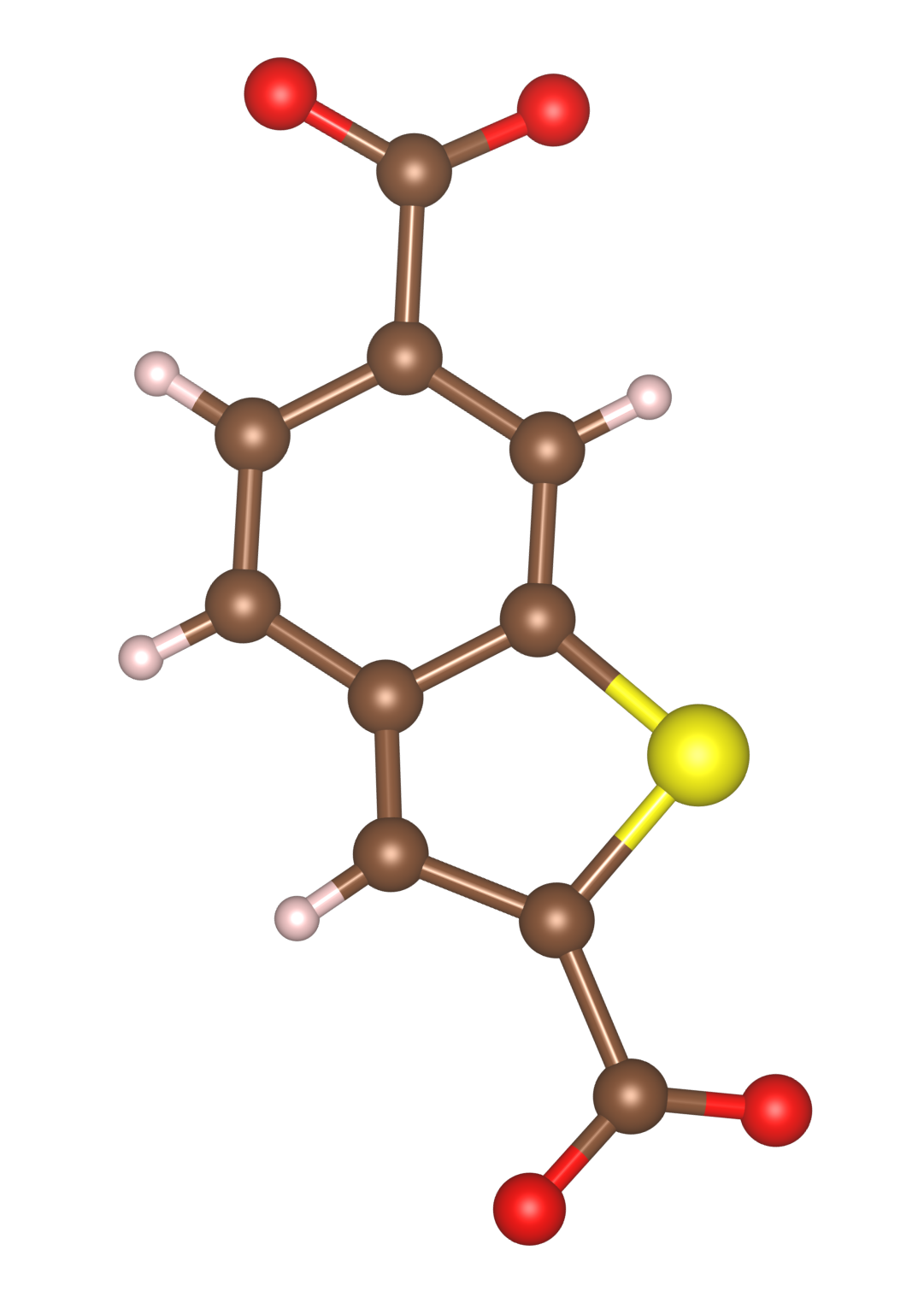}\\
		{\small (a) valid, $\Delta E \approx 0$~eV}
	\end{minipage}\hspace{0.04\textwidth}
	\begin{minipage}{0.35\textwidth}
		\centering
		\includegraphics[width=0.8\linewidth]{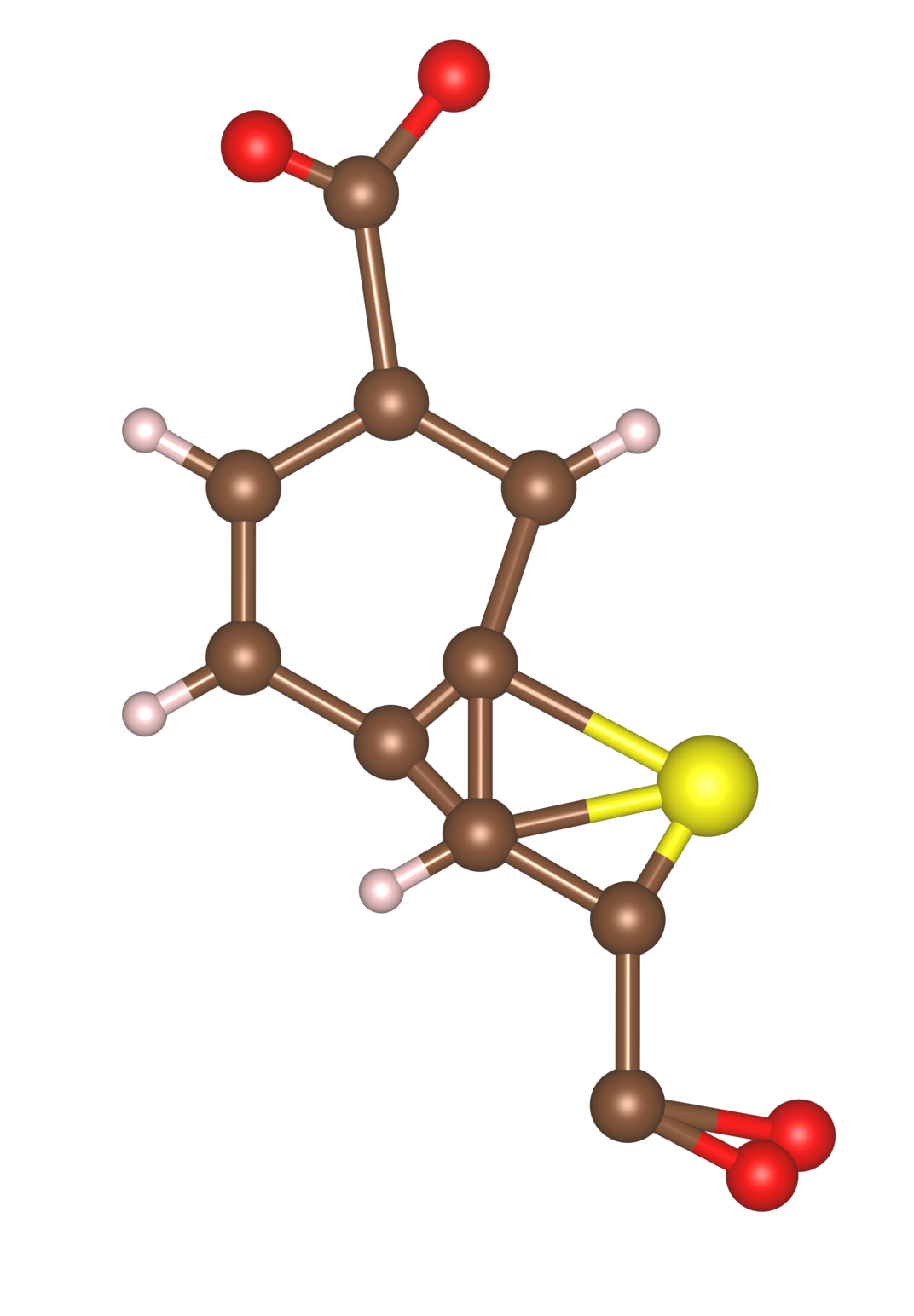}\\
		{\small (b) invalid, $\Delta E \approx 1.6 \times 10^4$~eV}
	\end{minipage}
	\caption{Representative matched-graph candidates for the organic linker
	(\ce{C10H4O4S}) of PICLAS. (a) A valid configuration ($\Delta E \approx 0$) is a
	physically reasonable conformation of the dynamically disordered linker.
	(b) An implausible candidate with high MLIP energy shows a distorted geometry
	and steric clashes.}
	\label{fig:piclas_candidates}
\end{figure}

When a disordered CIF yields multiple candidate configurations consistent with
the reference graph, the Inspector \& Editor uses the MLIP energy to separate
chemically plausible configurations from implausible ones. For the organic
linker (\ce{C10H4O4S}) of PICLAS, 1,728 candidate configurations were matched to
the reference graph. Their relative MLIP energies (\ce{MACE-OFF-24}) fall into
two well-separated groups (\Cref{fig:piclas_ehisto}). Only two configurations
lie at the energy minimum ($\Delta E \approx 0$), while the remaining 1,726 are
implausible, with relative energies reaching $\sim$34,000~eV from steric clashes.
The valid configurations correspond to physically reasonable conformations of the
dynamically disordered linker, whereas the high-energy candidates show distorted
geometries and atomic clashes (\Cref{fig:piclas_candidates}).

\newpage
\section{Details of LitMOF-DB}

\subsection{Database Composition and Correction Rates}

\Cref{tab:db_snapshot} summarizes the composition of LitMOF-DB by error category.
The CSD MOF subset contains 128,799 entries; 100 that failed to load from the CSD or
yielded an empty or unparsable framework were excluded, leaving 128,699 structures.
For each error category diagnosed over these structures, the table lists the number
identified, the number successfully corrected and exported to the released database, and
the resulting correction rate.
All counts are at the refcode level in the free-solvent-removed (FSR) state.

\begin{table}[!ht]
    \centering
    \caption{Per-category snapshot of LitMOF-DB, at the refcode level in the
    free-solvent-removed (FSR) state. ``Identified'' is the classifier output over the
    full CSD MOF subset; ``Final DB'' is the number of structures corrected and exported
    to the released database; ``Rate'' is Final\,/\,Identified. The 100 subset entries
    excluded for CSD-load failure or an empty or unparsable framework are not shown.
    Correct structures are retained without modification.}
    \label{tab:db_snapshot}
    \begin{tabular}{lrrr}
        \toprule
        Error category & Identified & Final DB & Rate \\
        \midrule
        Correct (no error) & 93{,}408 (72.6\%) & 93{,}408 & retained \\
        Hydrogen           & 22{,}920 (17.8\%) & 21{,}357 & 93.2\% \\
        Disorder           &  9{,}809 (7.6\%)  &  3{,}309 & 33.7\% \\
        Bond               &  2{,}562 (2.0\%)  &  2{,}345 & 91.5\% \\
        \midrule
        Total              & 128{,}699         & 120{,}419 & --- \\
        \bottomrule
    \end{tabular}
\end{table}

\subsection{Manual Validation of Corrections}

To assess the reliability of the corrections in LitMOF-DB, we manually validated 1,000
structures from the free-solvent-removed (FSR) set, stratified by the error type
diagnosed by LitMOF.
The set comprises 400 hydrogen-error, 200 disorder-error, and 200 bond-error structures,
together with 200 structures diagnosed as already correct.
For each structure, the corrected CIF was compared against the original publication and
the corresponding CSD entry, and judged valid when its connectivity and atomic
composition matched the reported structure.
The results are summarized in \Cref{tab:manual_validation}.

Corrections are valid for all bond-error and already-correct structures.
The invalid cases comprise 9 hydrogen corrections (out of 400, 97.8~\% valid) and 3
disorder corrections (out of 200, 98.5~\% valid). In the hydrogen cases, the
experimentally constrained heavy-atom positions force otherwise chemically reasonable
hydrogen placements into steric clashes, for which the MOF with refcode NORYUS
(\Cref{fig:noryus}) is a representative example.
The three invalid disorder cases follow the same pattern. They are valid as graph
representations but adopt implausible geometries, because the fixed heavy-atom positions
cannot assemble a chemically reasonable molecular structure, so the correction is
completed with a sound graph but an unsound geometry. Correcting them properly would
require remodeling the structure from scratch or performing a geometry optimization,
which would introduce hypothetical structure modeling and make the result a modeled
rather than an experimental MOF database, contrary to the construction principle of
LitMOF-DB.
The overall validity across the 1,000-structure set is 98.8~\%.

\begin{table}[!ht]
    \centering
    \caption{Manual validation of LitMOF corrections, stratified by error type
    (1,000 structures total). Validity is the fraction of corrected structures judged
    valid against the original publication and the corresponding CSD entry.}
    \label{tab:manual_validation}
    \begin{tabular}{l r l}
        \toprule
        Error type & Validated & Valid \\
        \midrule
        Hydrogen            & 400 & 391 (97.8\%) \\
        Disorder            & 200 & 197 (98.5\%) \\
        Bond                & 200 & 200 (100\%) \\
        Correct (no error)  & 200 & 200 (100\%) \\
        \midrule
        \textbf{Total}      & \textbf{1{,}000} & \textbf{988 (98.8\%)} \\
        \bottomrule
    \end{tabular}
\end{table}

\subsection{Correction Example with Impossible Case}
\begin{figure}[!htbp]
	\centering
	\includegraphics[width=0.9\textwidth]{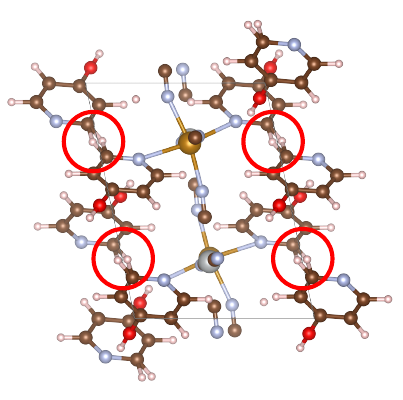}
	\caption{Corrected image of MOF NORYUS. The red circle indicates invalid hydrogen position.}
	\label{fig:noryus}
\end{figure}

\begin{figure}[!htbp]
	\centering
	\includegraphics[width=0.55\textwidth]{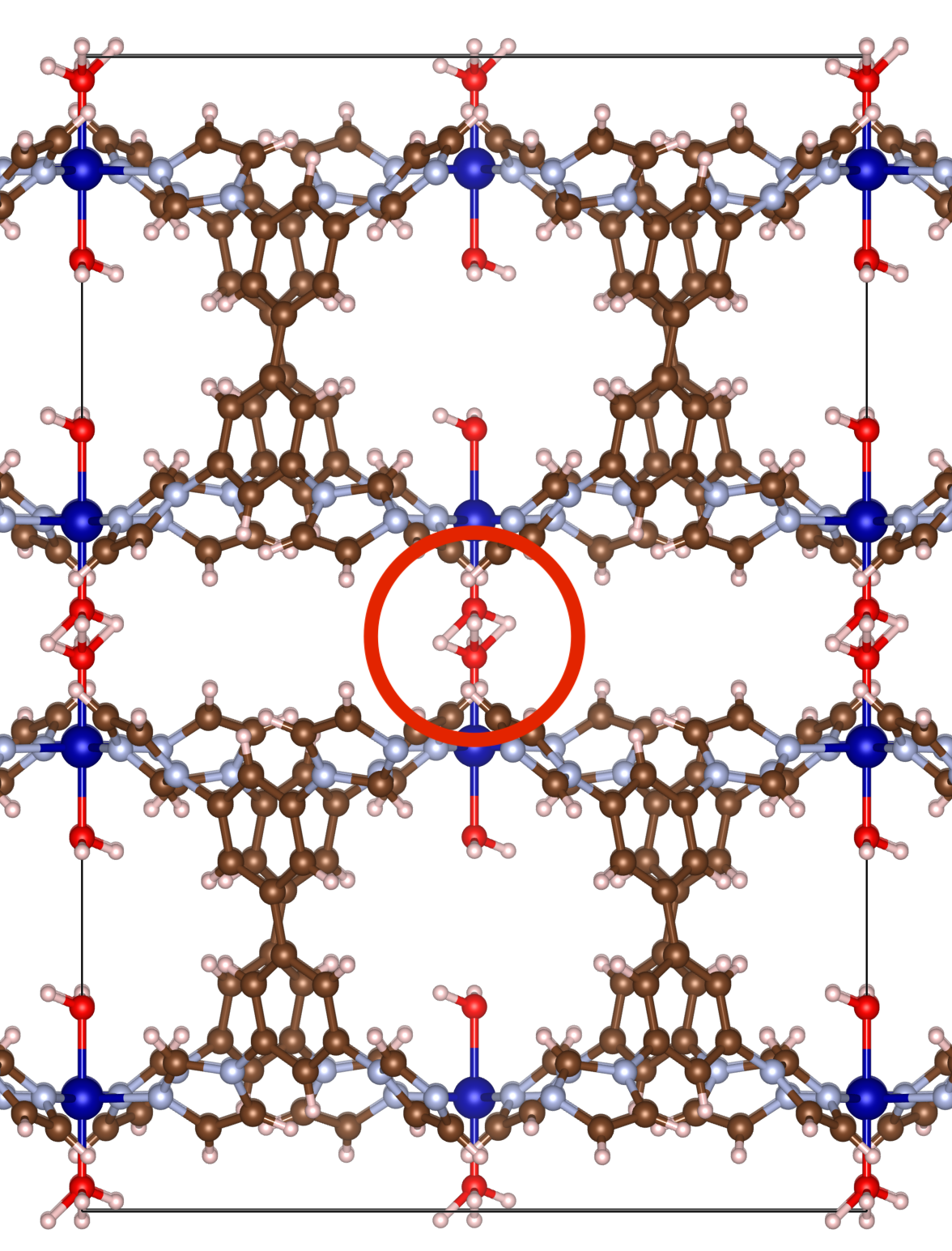}
	\caption{Corrected image of MOF AGAMAA01 (\ce{Co(H2O)2}(TIB)). The
	experimentally fixed oxygen positions of adjacent cobalt centres are too
	close, so completing the water molecules forces the added hydrogen atoms into
	an implausible geometry.}
	\label{fig:agamaa}
\end{figure}

As noted in the manual validation above, several of the hydrogen invalid cases arise
where a chemically reasonable hydrogen placement is geometrically irreconcilable with
the experimentally fixed heavy-atom positions. NORYUS and AGAMAA01 are
representative examples (\Cref{fig:noryus,fig:agamaa}).

AGAMAA01 has the structural formula \ce{Co(H2O)2}(TIB), where TIB is
3,3$'$,5,5$'$-tetra(1H-imidazol-1-yl)-1,1$'$-biphenyl. Each cobalt centre is
octahedrally coordinated by four nitrogen atoms from the TIB linkers and two
water oxygen atoms. The experimentally determined oxygen positions place
adjacent cobalt centres so close together that protonating these oxygens to
complete the water molecules forces the added hydrogen atoms into steric clash,
so no chemically valid hydrogen placement exists on the fixed heavy-atom
framework (\Cref{fig:agamaa}).

\newpage
\subsection{Perturbation-Recovery Benchmark}

To test the corrections against a known ground truth, we built a perturbation-recovery
benchmark for the two correction types that modify the structure, hydrogen and disorder.
In both cases we start from structures that LitMOF diagnoses as already correct, so the
original serves as ground truth, deliberately corrupt them, and test whether the
correction restores the original.

\begin{figure}[!ht]
\centering
\begin{minipage}{0.46\linewidth}\centering
\includegraphics[width=\linewidth]{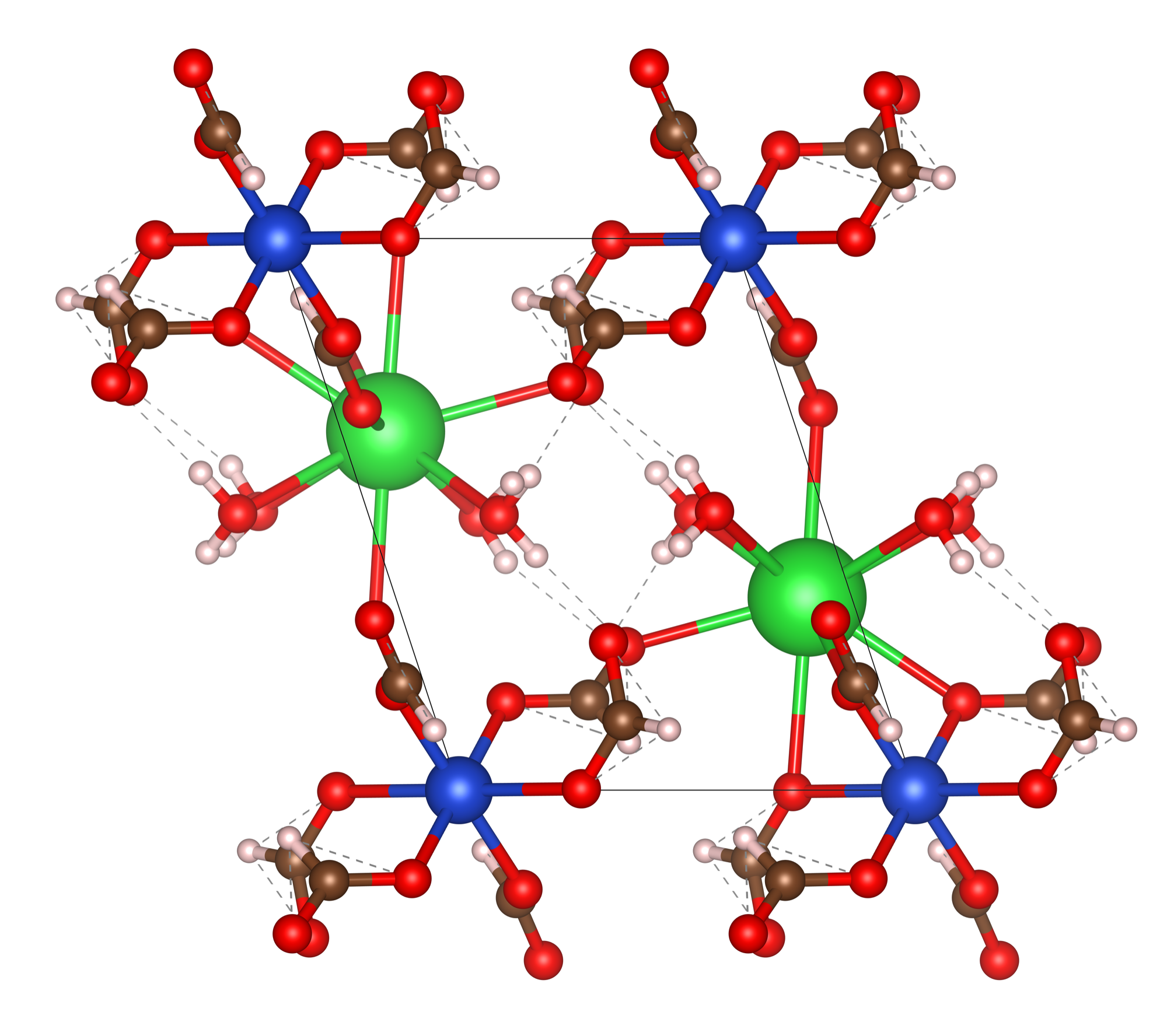}\\
(a) correct
\end{minipage}\hfill
\begin{minipage}{0.46\linewidth}\centering
\includegraphics[width=\linewidth]{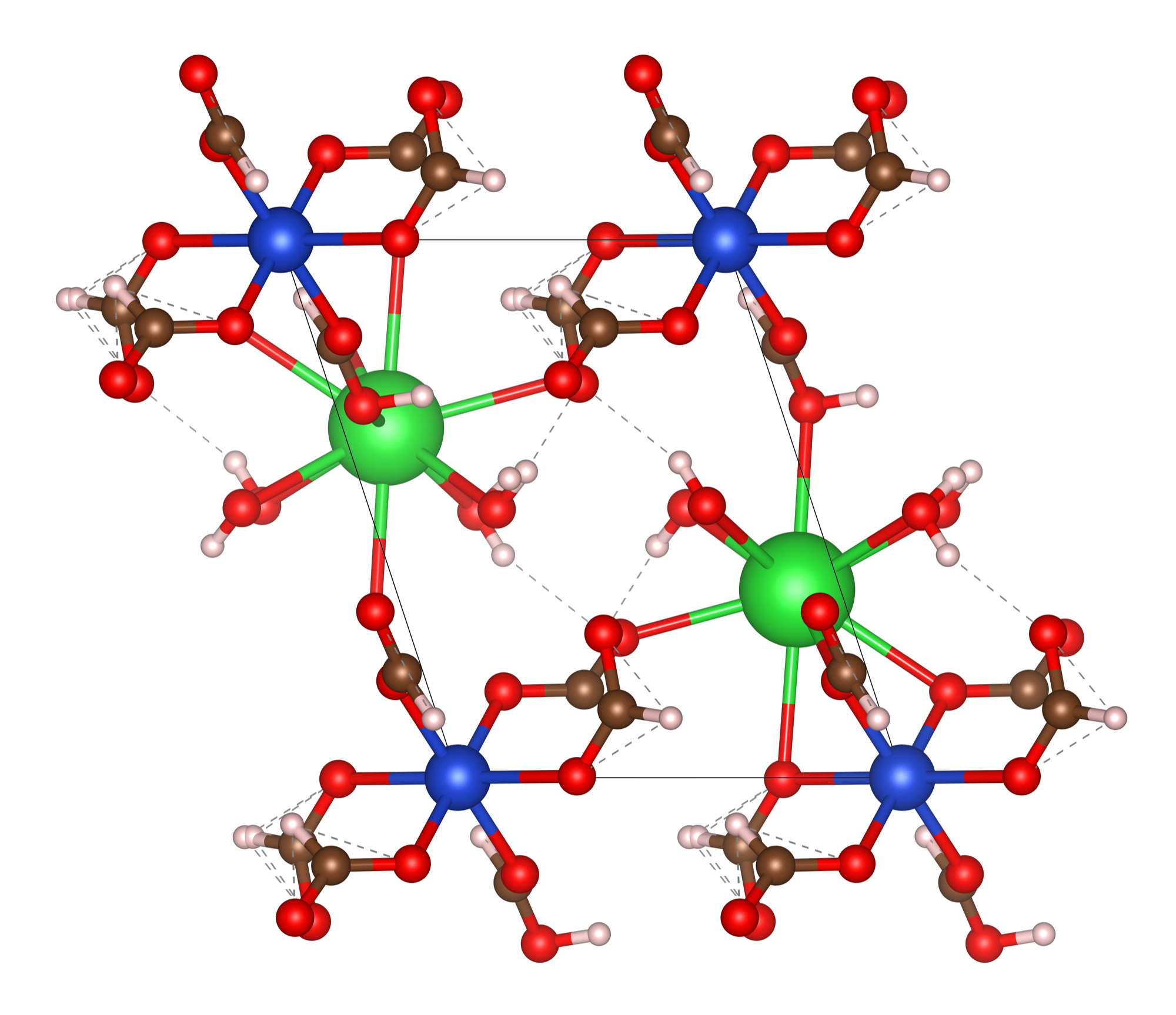}\\
(b) perturbed
\end{minipage}
\caption{Hydrogen perturbation used in the recovery benchmark, shown for refcode ABACUF01.
  (a) Correct structure with the known-valid hydrogen assignment (ground truth).
  (b) Perturbed structure after removing about 30\% of the hydrogen atoms and adding about
  10\% spurious hydrogens at random heavy-atom sites.}
\label{fig:hperturb}
\end{figure}

For hydrogen, we took 500 such structures and corrupted the hydrogen substructure of each
by removing about 30\% of the hydrogen atoms and adding about 10\% spurious hydrogens at
random heavy-atom sites (\Cref{fig:hperturb}). A structure was counted as recovered when
both the per-heavy-atom and the total hydrogen counts matched the original.

\begin{figure}[!ht]
\centering
\begin{minipage}{0.31\linewidth}\centering
\includegraphics[width=\linewidth]{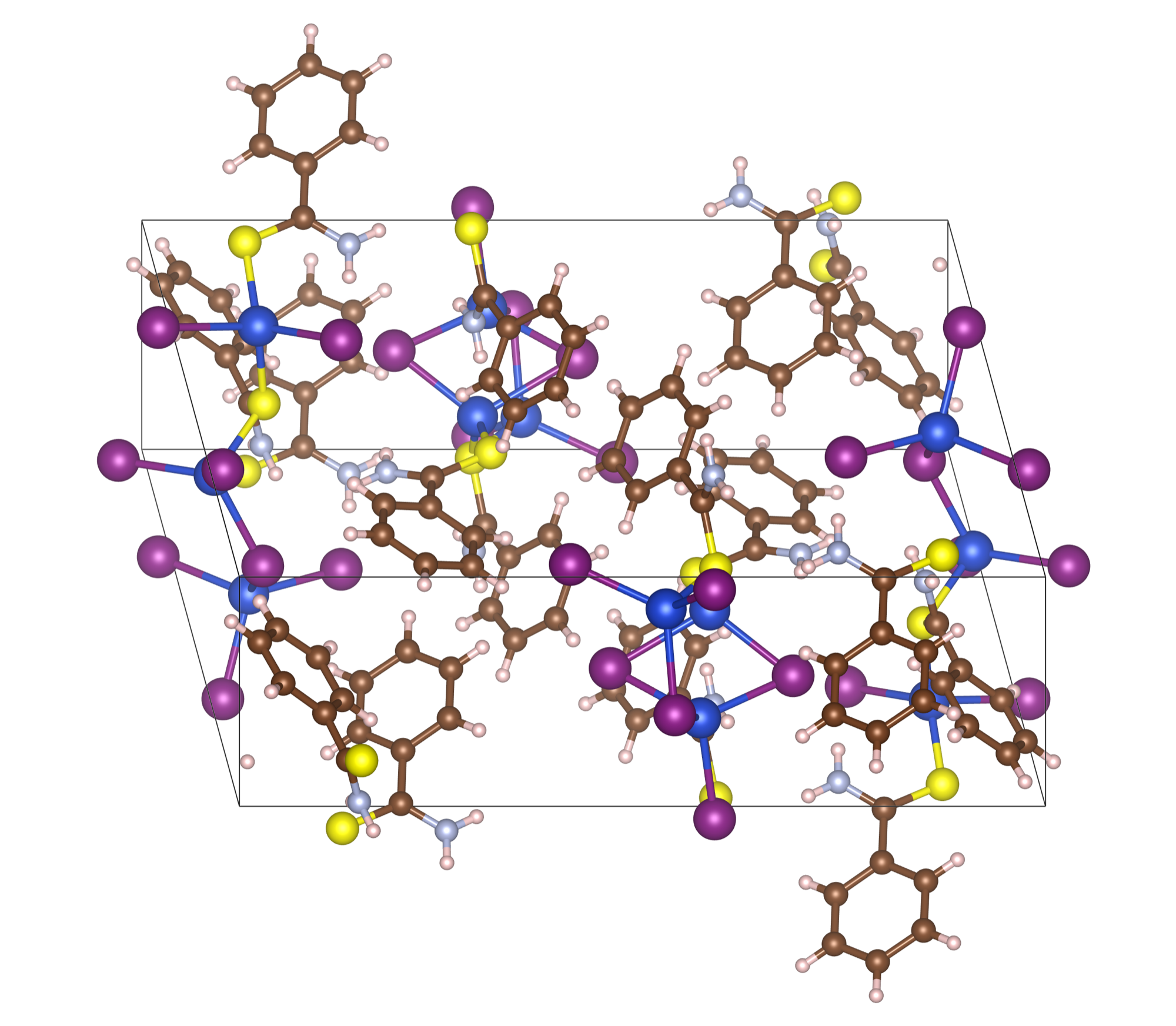}\\
(a) correct
\end{minipage}\hfill
\begin{minipage}{0.31\linewidth}\centering
\includegraphics[width=\linewidth]{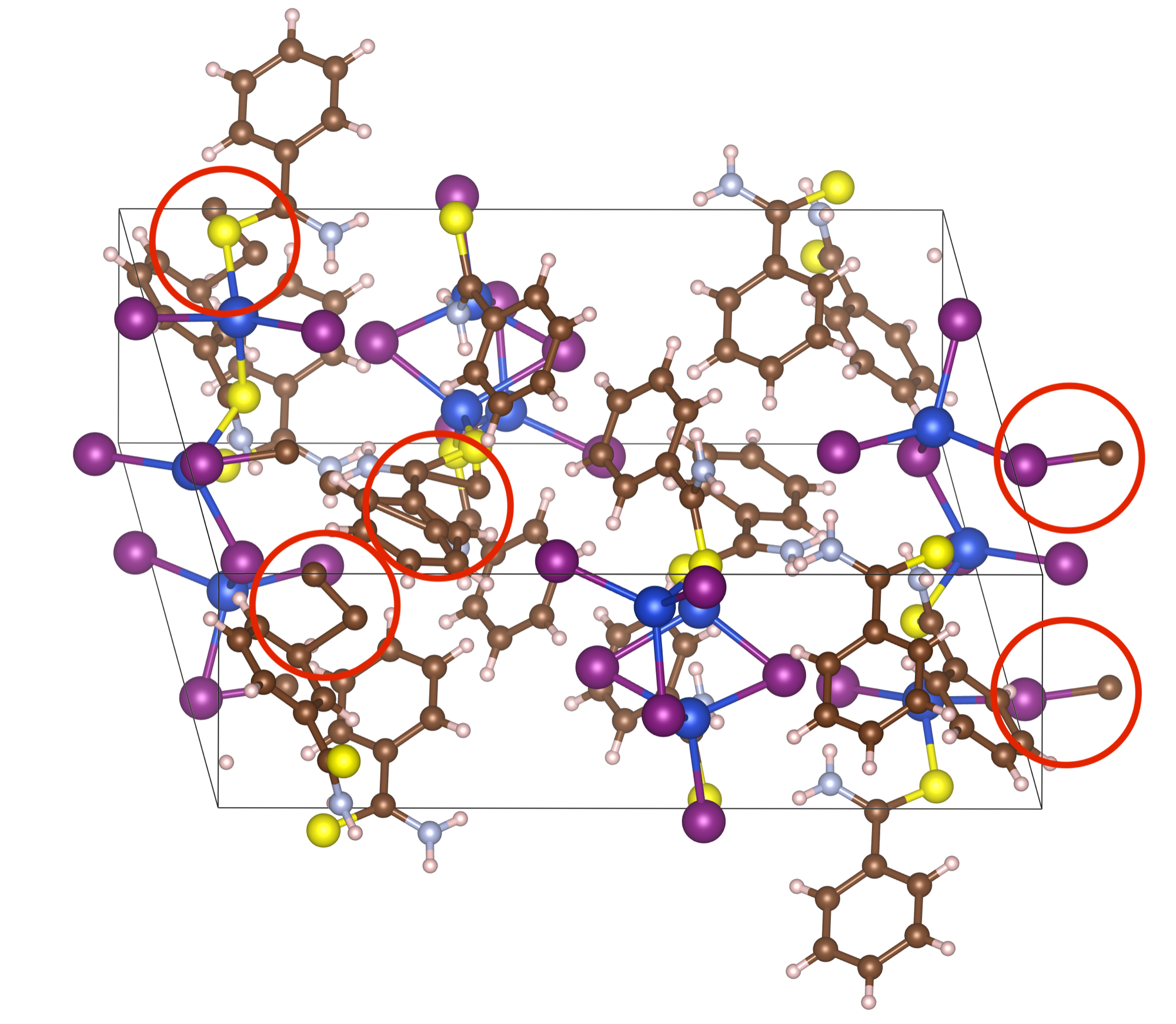}\\
(b) jitter
\end{minipage}\hfill
\begin{minipage}{0.31\linewidth}\centering
\includegraphics[width=\linewidth]{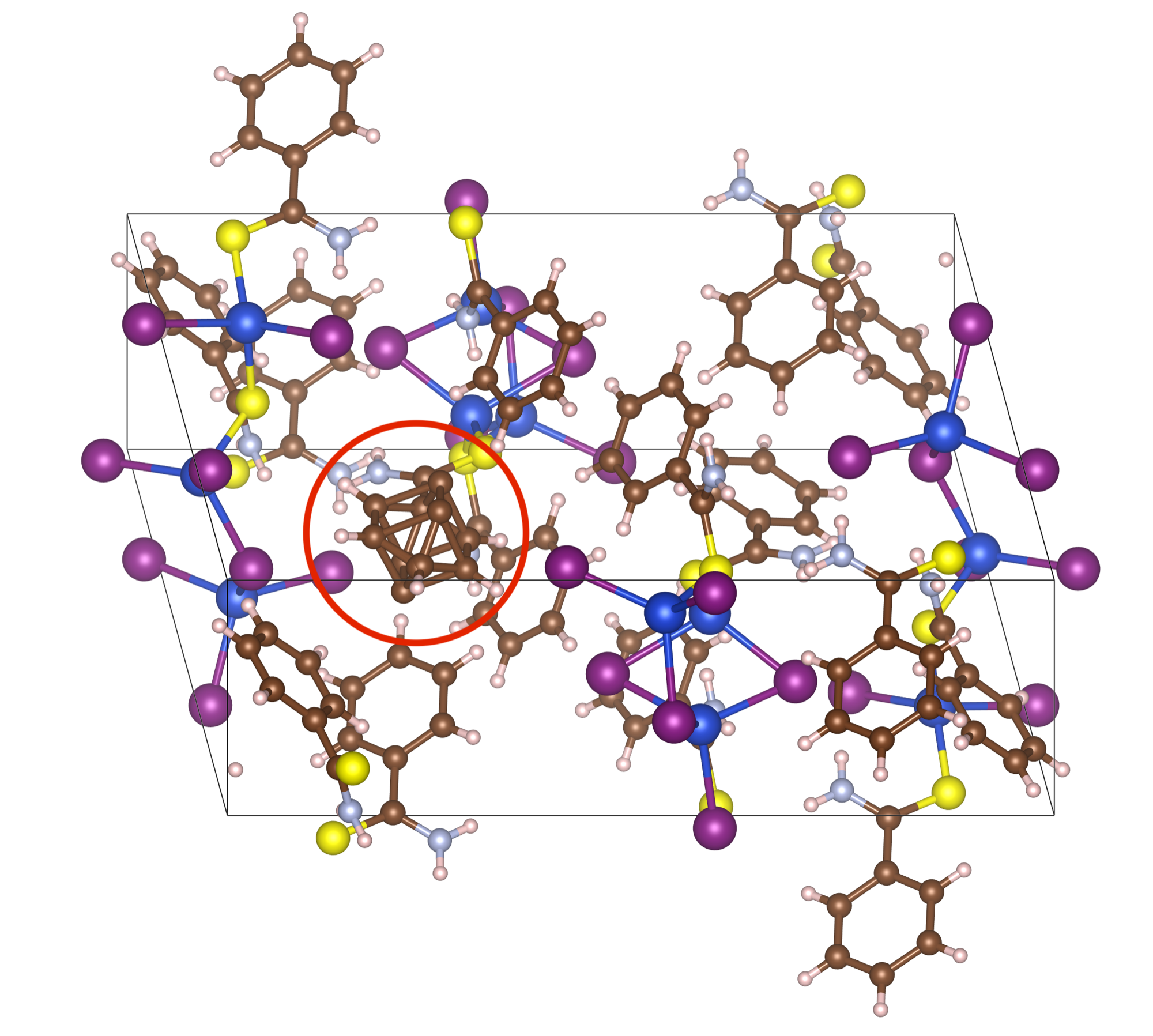}\\
(c) rotation
\end{minipage}
\caption{Disorder perturbations used in the recovery benchmark, shown for refcode
  TIYZOT01. (a) Correct structure (ground truth). (b) Coordinate jitter applied to a
  fragment. (c) Rotation of a fragment about its anchor. The red circle marks the
  perturbed region.}
\label{fig:dperturb}
\end{figure}

For disorder, we injected synthetic disorder into 100 structures by duplicating a terminal
fragment of a linker at occupancy 0.5, bonded to the same anchor atom but placed at a
perturbed position, which reproduces a two-orientation disorder site (\Cref{fig:dperturb}).
Two perturbation types were used, a coordinate jitter that displaces the duplicated atoms
by Gaussian noise of 0.8~\AA\ standard deviation, and a fragment rotation that rotates the
duplicate by $90^\circ$ about its anchor bond. A structure was counted as recovered when
the species count and the structure-validity checks matched the original.

\begin{table}[!ht]
    \centering
    \caption{Perturbation-recovery test. Known-correct structures were deliberately
    corrupted and passed through the corresponding LitMOF correction. ``Recovered'' is the
    number whose structure was restored to the original.}
    \label{tab:perturb_recovery}
    \begin{tabular}{lrr}
        \toprule
        Perturbation type & Tested & Recovered \\
        \midrule
        Hydrogen             & 500 & 476 (95.2\%) \\
        Disorder (jitter)    & 100 & 88 (88\%) \\
        Disorder (rotation)  & 100 & 87 (87\%) \\
        \bottomrule
    \end{tabular}
\end{table}

Both corrections recover the original structure at high rates when the ground truth is
known, 95.2\% for hydrogen and 88\% and 87\% for the two disorder perturbations
(\Cref{tab:perturb_recovery}). The
disorder rate here is well above the 33.7\% disorder-correction rate of LitMOF-DB because
the benchmark starts from a reference graph already known to be correct and applies
relatively simple synthetic perturbations, whereas the database rate reflects the more
complex disorder present in experimental CIFs.

\newpage
\section{Evaluation with MOFChecker, MOFClassifier, and MOSAEC}

To assess whether the corrections in LitMOF-DB are objectively beneficial, we re-scored
every structure with three independent, third-party validators.
MOFChecker applies rule-based coordination checks\cite{Jablonka_mofchecker_2023}, MOFClassifier
provides a machine-learned crystal-likeness score\cite{zhao2025mofclassifier}, and MOSAEC
evaluates metal oxidation-state consistency\cite{white2025high}.
None of the three is the reference against which our correction is optimized, so this is
an external check rather than a self-evaluation.
We compare each structure in LitMOF-DB (FSR) before and after LitMOF correction, using the
same refcode in both states. The before structure is the CSD MOF entry with only the free
solvent removed, and the after structure is its LitMOF-corrected form. The comparison is
stratified by the diagnosed error category and the framework charge state (\Cref{tab:eval}).

\begin{table}[!ht]
    \centering
    \small
    \caption{External validation of LitMOF-DB by error category and framework charge.
    MOFChecker and MOSAEC values are the fraction of structures flagged clean (\%), and
    MOFClassifier is the mean crystal-likeness score; $\uparrow$ indicates that higher is
    better. Values are computed on a paired basis over the structures present in
    LitMOF-DB (total 120,419). For the Correct and Bond categories the structure is
    unchanged with respect to these validators (no atoms are repositioned, and bond-order
    corrections do not alter the coordination- or geometry-based scores), so a single
    value is shown; for Hydrogen and Disorder the values are given as
    before~$\to$~after.}
    \label{tab:eval}
    \begin{tabular*}{\textwidth}{@{\extracolsep{\fill}}llrccc}
        \toprule
        Category & Charge & $n$ & MOFChecker~$\uparrow$ & MOSAEC~$\uparrow$ & MOFClassifier~$\uparrow$ \\
        \midrule
        \multirow{2}{*}{Correct}  & neutral & 77{,}848 & 78.2 & 92.4 & 0.671 \\
                                  & charged & 15{,}560 & 76.9 & 20.2 & 0.418 \\
        \midrule
        \multirow{2}{*}{Bond}     & neutral &  2{,}105 & 77.7 & 85.7 & 0.572 \\
                                  & charged &    240 & 82.5 & 15.0 & 0.473 \\
        \midrule
        \multirow{2}{*}{Hydrogen} & neutral & 17{,}697 & 25.4~$\to$~\textbf{75.8} & 15.8~$\to$~\textbf{79.7} & 0.216~$\to$~\textbf{0.614} \\
                                  & charged &  3{,}660 & 29.0~$\to$~\textbf{72.7} & 11.8~$\to$~\textbf{17.0} & 0.124~$\to$~\textbf{0.433} \\
        \midrule
        \multirow{2}{*}{Disorder} & neutral &  2{,}779 & 22.4~$\to$~\textbf{67.0} & 23.3~$\to$~\textbf{68.0} & 0.197~$\to$~\textbf{0.580} \\
                                  & charged &    530 & 20.2~$\to$~\textbf{75.1} &  9.1~$\to$~\textbf{16.4} & 0.117~$\to$~\textbf{0.400} \\
        \midrule
        \textbf{Total} &          & 120{,}419 & 67.2~$\to$~\textbf{77.2} & 67.1~$\to$~\textbf{77.7} & 0.540~$\to$~\textbf{0.617} \\
        \bottomrule
    \end{tabular*}
\end{table}

The Correct and Bond categories are unchanged on all three validators.
Correct structures are retained without modification, and although Bond corrections rewrite
the bond records in the CIF, these validators do not read that information; they re-derive
connectivity from interatomic distances (a distance-based adjacency matrix), so writing the
correct bond information into the CIF has no effect on their scores.
The Hydrogen and Disorder categories, which alter the atomic content, improve
substantially on every validator for neutral frameworks.
MOSAEC stays near-flat on the charged subset (for example, Hydrogen charged
15.8~$\to$~17.0\%); this is by construction, because a free-solvent-removed framework
with its counterion stripped is intentionally net-charged, and MOSAEC flags that
imbalance regardless of how well the framework itself is corrected.
The charged subset should therefore be judged by MOFChecker and MOFClassifier, which both
improve clearly.

Overall, the corrections in LitMOF-DB produce a substantial and consistent improvement
across these independent external metrics, which provides further evidence for the
correctness of the database.
The manual validation of 1,000 structures (\Cref{tab:manual_validation}; 98.8~\% valid),
however, places the true reliability of LitMOF-DB well above what these automated metrics
indicate.
The gap arises because these tools depend on a fixed algorithm or a machine-learning
model, and their verdicts are not fully accurate.
What matters most here is recall rather than precision, since a tool may be precise and
still discard a substantial number of structures that are in fact correct.
Robust validation therefore requires reconciling several independent sources, as in the
construction of LitMOF-DB, where the CSD metadata, the CIF graph, and the primary
literature are cross-checked, a reconciliation made possible by the LLM.
The practical cost of low recall, namely that genuinely usable structures are missed, is
demonstrated by the \ce{CO2} direct air capture screening in the main text (Section~2.5),
where many high-performing candidates are absent from databases built on single-criterion
tools.

\newpage
\section{Comparison with Other MOF Databases}

\subsection{Different Categorization of Solvent Removal}

The databases compared here organize solvent-removed structures under different category
schemes, which must be reconciled before any cross-database comparison.
In all cases the free solvent is removed first; the databases then differ in how the
resulting structures are labeled and stored.

For a MOF that contains no bound solvent (that is, only free solvent or no solvent at
all), free-solvent removal already yields the final structure.
CoRE MOF DB stores this structure redundantly under both the FSR and ASR categories;
MOSAEC-DB stores it only under the full (fully activated) category; LitMOF-DB
stores it only under FSR.

For a MOF that contains bound solvent, two structures are generated.
Removing only the free solvent gives a structure that CoRE MOF DB and LitMOF-DB store under
FSR, whereas MOSAEC-DB assigns it to the partial category.
Removing the bound solvent as well gives a fully activated structure, which CoRE MOF DB
stores under ASR, LitMOF-DB under BSR, and MOSAEC-DB under full.
These correspondences are summarized in \Cref{tab:solvent_categorization}.

\begin{table}[!ht]
    \centering
    \caption{Category under which each database stores a structure after solvent removal,
    for MOFs with and without bound solvent. Free solvent is removed in all cases; the
    schemes differ in the treatment of bound solvent. CoRE MOF DB stores a bound-solvent-free
    structure redundantly under both FSR and ASR.}
    \label{tab:solvent_categorization}
    \begin{tabularx}{\textwidth}{@{} X c c c @{}}
        \toprule
        Structure & CoRE MOF DB & MOSAEC-DB & LitMOF-DB \\
        \midrule
        No bound solvent (free solvent removed) & FSR and ASR (duplicated) & full & FSR \\
        Bound solvent, free solvent removed & FSR & partial & FSR \\
        Bound solvent, free and bound solvent removed & ASR & full & BSR \\
        \bottomrule
    \end{tabularx}
\end{table}

\subsection{CoRE MOF DB}

We downloaded the CoRE MOF 2024 v1.1 database\cite{zhao2025core, zhao_2025_15055758} and used it as the
reference for comparison with LitMOF-DB.
The database is divided into computation-ready (CR) and not-computation-ready (NCR)
groups, and each group is provided under three solvent-removal protocols: all-solvent
removed (ASR), free-solvent removed (FSR), and ion-removed (ION).
Within each protocol, entries are further labeled as CSD\_modified, CSD\_unmodified, or
SI, where the first two indicate whether the structure was altered relative to its CSD
entry during database construction.
The composition of the database we obtained is summarized in \Cref{tab:coremof_composition}.
The counts differ slightly from those reported in the original CoRE MOF
publication\cite{zhao2025core}; this discrepancy arises from the difference in the
underlying CSD version used to construct the database.

\begin{table}[!ht]
    \centering
    \caption{Composition of the CoRE MOF DB\cite{zhao2025core, zhao_2025_15055758} used in
    this work, classified by computation-readiness (CR/NCR), solvent-removal protocol
    (ASR/FSR/ION), and curation label (CSD\_modified/CSD\_unmodified/SI). The
    CSD\_modified and CSD\_unmodified labels indicate whether the structure was altered
    relative to its CSD entry during database construction, and are unrelated to the
    structural corrections applied by LitMOF. SI-labeled structures,
    which originate from supporting information and lack a CSD refcode, are excluded
    from the comparison with LitMOF-DB, as LitMOF is restricted to entries with a CSD
    refcode. Parenthetical values in the ASR rows denote the number of unique ASR
    structures, defined as those differing in elemental composition from their FSR
    counterpart or lacking an FSR counterpart; the remainder are duplicates of FSR
    structures.}
    \label{tab:coremof_composition}
    \begin{tabular*}{\textwidth}{@{\extracolsep{\fill}} l l r r r r}
        \toprule
        Group & Protocol & CSD\_modified & CSD\_unmodified & SI & Total \\
        \midrule
        \multirow{4}{*}{CR}
            & ASR & 5{,}591 (3{,}113) & 1{,}893 (2) & 1{,}372 (647) & 8{,}856 (3{,}762) \\
            & FSR & 3{,}786 & 2{,}656 & 1{,}192 & 7{,}634 \\
            & ION &   458 &   152 &   100 &   710 \\
        \cmidrule{2-6}
            & \textbf{Total} & & & & \textbf{17{,}200} \\
        \midrule
        \multirow{4}{*}{NCR}
            & ASR & 5{,}736 (3{,}500) & 2{,}352 (27) & 2{,}593 (1{,}058) & 10{,}681 (4{,}585) \\
            & FSR & 4{,}209 & 4{,}738 & 2{,}811 & 11{,}758 \\
            & ION &   496 &   465 &   232 &  1{,}193 \\
        \cmidrule{2-6}
            & \textbf{Total} & & & & \textbf{23{,}632} \\
        \bottomrule
    \end{tabular*}
\end{table}

Unlike MOSAEC-DB and LitMOF-DB, the CoRE MOF DB can contain the same CIF in both
the ASR and FSR sets: when a free-solvent-removed (FSR) structure contains no bound
solvent, the resulting structure is identical to the all-solvent-removed (ASR) structure
and is duplicated across the two sets.
This duplication is substantial: of the 8{,}856 CR ASR entries, only 4{,}488 are
genuinely unique (i.e.\ a bound solvent was actually removed, or no FSR counterpart
exists), and the corresponding figure for the 10{,}681 NCR ASR entries is 4{,}577
(\Cref{tab:coremof_composition}).
To avoid this ambiguity, we adopt a distinct naming convention in LitMOF and refer to the
corresponding structures as bound-solvent removed (BSR).

Because LitMOF begins from MOF entries that carry a CSD refcode, the SI-labeled
structures in the CoRE MOF DB, which originate from the supporting information of
publications and do not have a corresponding CSD refcode, fall outside the scope of
LitMOF and were therefore not included in the comparison.

We compared LitMOF-DB against the CoRE MOF DB separately for each solvent-removal
treatment (FSR, ION, and BSR; the CoRE MOF DB ASR set is used as the BSR counterpart, see
\Cref{tab:solvent_categorization}).
For every refcode present in both databases, the unit-cell composition was compared after
expanding all crystallographic symmetry operations with ASE, so that structures are
judged identical whenever they describe the same crystal regardless of cell convention.
Differences are classified as \texttt{diff\_H} (the structures differ only in hydrogen,
that is, a protonation or hydrogen-convention difference) or \texttt{diff\_heavy} (a
genuine difference in the heavy-atom content, such as a removed solvent or a modified
framework).

Two quantities are reported.
First, the \emph{NCR recovery rate}: the fraction of CoRE MOF DB not-computation-ready (NCR)
entries that LitMOF retains as valid, computation-ready structures
(\Cref{tab:coremof_ncr}).
Second, the \emph{CR correction rate}: the fraction of CoRE MOF DB computation-ready (CR)
entries whose composition LitMOF changed (\Cref{tab:coremof_cr}).
A changed composition for a CR entry means that the structure CoRE MOF DB labeled as
computation-ready in fact required correction, so this rate quantifies the false-positive
rate of the CoRE MOF DB CR label.

\begin{table}[!ht]
    \centering
    \caption{Recovery of CoRE MOF DB not-computation-ready (NCR) entries by LitMOF, per
    solvent-removal treatment (columns). The Total row gives the overall recovery rate as
    recovered/NCR-total: the number of NCR entries that LitMOF retains as computation-ready
    over the CoRE MOF DB NCR total for that treatment. The same / \texttt{diff\_H} /
    \texttt{diff\_heavy} rows break the recovered structures down by how they compare with
    the CoRE MOF DB composition (identical / differing only in hydrogen / differing in heavy
    atoms); these percentages are shares of the recovered set for that treatment and
    therefore sum to 100\%. The BSR denominator is the set of unique ASR refcodes (those
    whose composition differs from their FSR counterpart).}
    \label{tab:coremof_ncr}
    \begin{tabular*}{\textwidth}{@{\extracolsep{\fill}} l r r r}
        \toprule
        Comparison & BSR & FSR & ION \\
        \midrule
        same                &   863 (40.3\%) & 1{,}445 (22.0\%) & 182 (31.3\%) \\
        \texttt{diff\_H}    &   273 (12.7\%) & 1{,}854 (28.3\%) & 135 (23.2\%) \\
        \texttt{diff\_heavy}& 1{,}006 (47.0\%) & 3{,}254 (49.7\%) & 265 (45.5\%) \\
        \midrule
        Total (recovered)   & 2{,}142/3{,}527 (60.7\%) & 6{,}553/8{,}947 (73.2\%) & 582/961 (60.6\%) \\
        \bottomrule
    \end{tabular*}
\end{table}

\begin{table}[!ht]
    \centering
    \caption{Correction of CoRE MOF DB computation-ready (CR) entries by LitMOF, per
    solvent-removal treatment (columns). The Total row gives the correction rate as
    corrected/CR-total: the number of CR entries whose composition LitMOF changed over the
    CoRE MOF DB CR total for that treatment; because CoRE MOF DB had labeled these entries
    computation-ready, this is the false-positive rate of the CoRE MOF DB CR label. The
    \texttt{diff\_H} / \texttt{diff\_heavy} rows break the corrected structures down by the
    type of change; these percentages are shares of the corrected set for that treatment
    and therefore sum to 100\%. The BSR denominator is the set of unique ASR refcodes.}
    \label{tab:coremof_cr}
    \begin{tabular*}{\textwidth}{@{\extracolsep{\fill}} l r r r}
        \toprule
        Comparison & BSR & FSR & ION \\
        \midrule
        \texttt{diff\_H}    & 177 (37.5\%) & 629 (76.2\%) &  42 (28.2\%) \\
        \texttt{diff\_heavy}& 295 (62.5\%) & 196 (23.8\%) & 107 (71.8\%) \\
        \midrule
        Total (corrected)   & 472/3{,}115 (15.2\%) & 825/6{,}442 (12.8\%) & 149/610 (24.4\%) \\
        \bottomrule
    \end{tabular*}
\end{table}

LitMOF recovers the majority of NCR entries across all treatments (73.2~\% for FSR), and
identifies a substantial fraction of CoRE MOF DB CR entries as requiring correction
(12.8~\%, 15.2~\%, and 24.4~\% for FSR, BSR, and ION, respectively), demonstrating that a
non-negligible portion of the structures CoRE MOF DB treats as computation-ready still carry
underlying structural errors.

\subsection{MOSAEC-DB}

We compared LitMOF-DB with MOSAEC-DB at the refcode level
(\Cref{tab:mosaec_overlap}).
Both databases are built from the CSD, and the two structure sets overlap
substantially, with 83,633 shared refcodes.

\begin{table}[!ht]
    \centering
    \caption{Refcode-level overlap between LitMOF-DB and MOSAEC-DB.}
    \label{tab:mosaec_overlap}
    \begin{tabular}{lr}
        \toprule
        Overlap category & Refcodes \\
        \midrule
        Present in both LitMOF-DB and MOSAEC-DB & 83{,}633 \\
        LitMOF-DB only & 36{,}786 \\
        MOSAEC-DB only &  8{,}110 \\
        \bottomrule
    \end{tabular}
\end{table}

For the shared refcodes, we compared the corrected structures for each
solvent-removal treatment (\Cref{tab:mosaec_category}).
The two workflows produce identical structures (the same composition including
hydrogen and a coordinate RMSD below 0.1~\AA) in 95.4~\% (FSR) and 95.2~\% (BSR)
of the matched cases, confirming that they agree where they overlap.
The ION treatment retains charged non-framework components and has no MOSAEC-DB
counterpart.

\begin{table}[!ht]
    \centering
    \caption{Category-wise comparison of LitMOF-DB and MOSAEC-DB. ``In MOSAEC-DB''
    is the number of matched refcodes; ``Identical'' is the fraction of matched
    pairs with the same composition (including hydrogen) and a coordinate RMSD
    below 0.1~\AA. The ION treatment has no MOSAEC-DB counterpart.}
    \label{tab:mosaec_category}
    \begin{tabular}{lrrr}
        \toprule
        Category & Total & In MOSAEC-DB & Identical (\%) \\
        \midrule
        FSR (free-solvent removed)        & 120{,}419 & 76{,}493 & 95.4 \\
        BSR (bound-solvent removed)       &  52{,}156 & 37{,}043 & 95.2 \\
        ION (charged components retained) &  16{,}992 &        0 & n/a  \\
        \bottomrule
    \end{tabular}
\end{table}

The 36,786 refcodes present only in LitMOF-DB are structures that MOSAEC-DB
discarded as invalid under its oxidation-state criterion but that LitMOF was
able to correct and validate.
Of the 8,110 refcodes present only in MOSAEC-DB, the majority fall outside
LitMOF's processing scope (\Cref{tab:mosaec_only}): 6,431 originate from CSD
entries outside the \texttt{Subsets.MOF} collection, and the remaining 1,679
are within the MOF subset but were dropped during LitMOF processing,
predominantly disorder-rebuild failures.

\begin{table}[!ht]
    \centering
    \caption{Breakdown of refcodes present in MOSAEC-DB but not in LitMOF-DB.
    The majority fall outside LitMOF's processing scope.}
    \label{tab:mosaec_only}
    \begin{tabular}{lr}
        \toprule
        Reason for MOSAEC-DB-only refcodes & Refcodes \\
        \midrule
        Outside CSD \texttt{Subsets.MOF} (outside LitMOF scope) & 6{,}431 \\
        Within CSD \texttt{Subsets.MOF} but dropped during LitMOF processing & 1{,}679 \\
        \midrule
        Total & 8{,}110 \\
        \bottomrule
    \end{tabular}
\end{table}

\newpage
\section{Supplementary Figures and Tables}

\begin{figure}[!ht]
	\centering
	\includegraphics[width=0.8\textwidth]{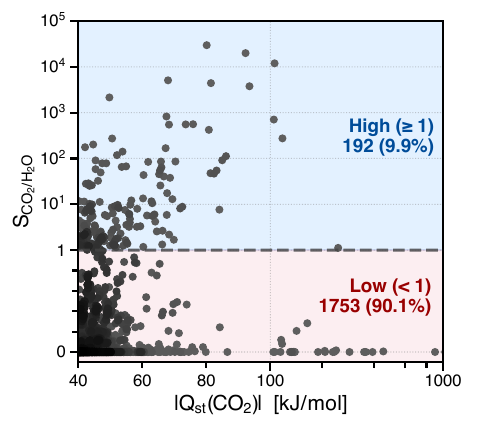}
	\caption{Heat of adsorption of \ce{CO2} vs selectivity of \ce{CO2} over \ce{H2O} plot on original MOF structures.}
	\label{fig:selectivity_original}
\end{figure}

\begin{figure}[!ht]
	\centering
	\includegraphics[width=0.8\textwidth]{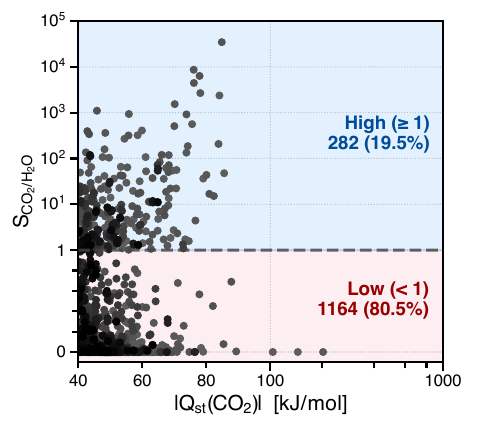}
	\caption{Heat of adsorption of \ce{CO2} vs selectivity of \ce{CO2} over \ce{H2O} plot on corrected MOF structures.}
	\label{fig:selectivity_corrected}
\end{figure}

\begin{table}[!ht]
    \centering
    \caption{Presence of the 236 DAC candidates recovered only after LitMOF
    correction in the computation-ready content of existing MOF databases.
    A candidate is counted as present when its refcode appears in the
    corresponding database. The majority are absent from both, which indicates
    that these candidates are not reachable by a screen based only on existing
    computation-ready databases.}
    \label{tab:dac_provenance}
    \begin{tabular}{lrr}
        \toprule
        External database & Present & Absent \\
        \midrule
        CoRE MOF DB & 61 & 175 \\
        MOSAEC-DB   & 80 & 156 \\
        \bottomrule
    \end{tabular}
\end{table}

\subsection{Classical Force-Field and GCMC Validation of DAC Screening}

\begin{figure}[!ht]
	\centering
	\includegraphics[width=0.7\textwidth]{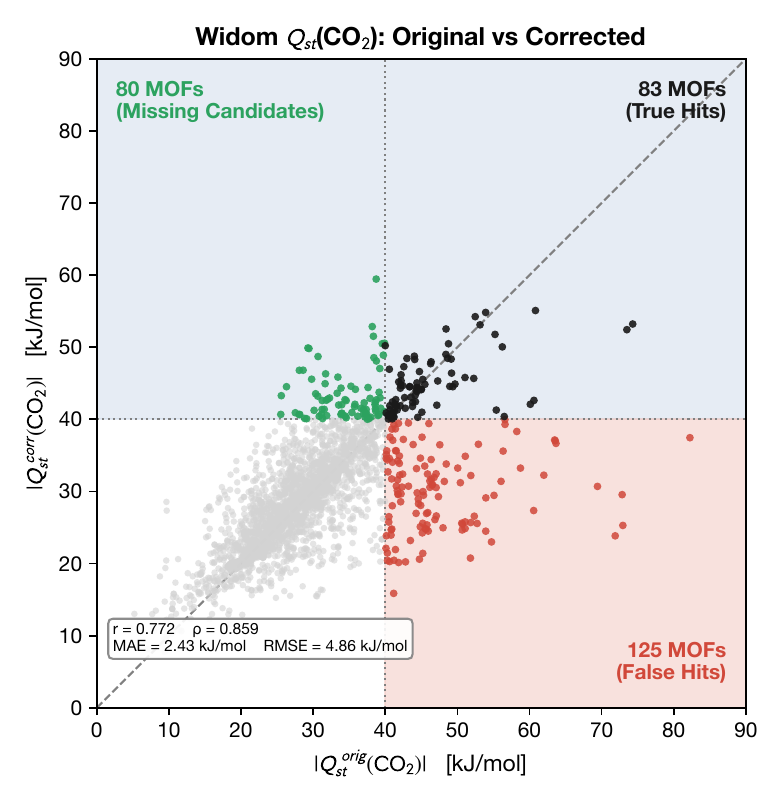}
	\caption{\ce{CO2} heat of adsorption from classical force-field Widom
	insertion for original (as-is, FSR) vs.\ LitMOF-corrected structures.
	Quadrant counts use the $|Q_\mathrm{st}(\ce{CO2})| > 40~\mathrm{kJ~mol^{-1}}$
	threshold.}
	\label{fig:widom_ff_co2}
\end{figure}

\begin{figure}[!ht]
	\centering
	\includegraphics[width=0.7\textwidth]{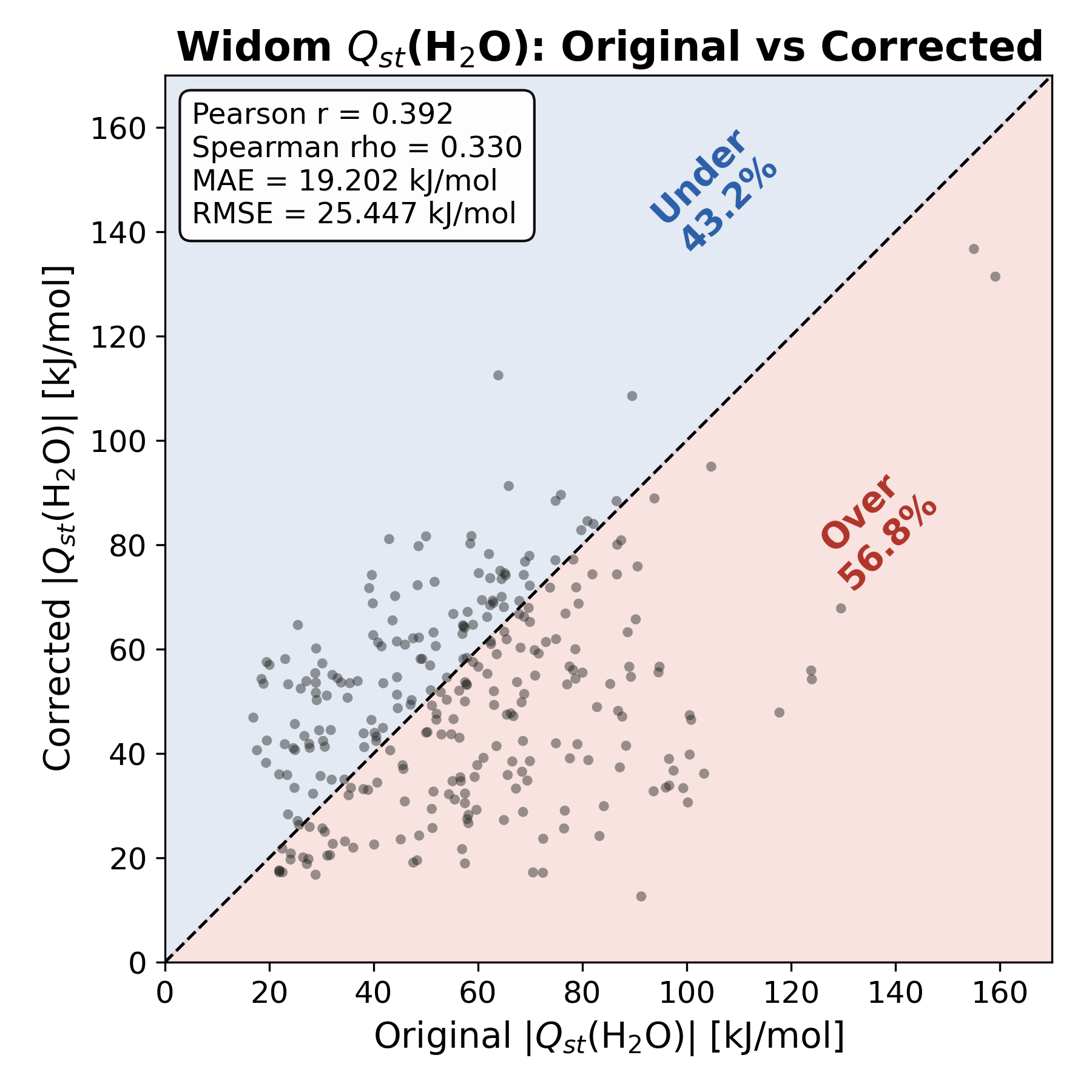}
	\caption{\ce{H2O} heat of adsorption from classical force-field Widom
	insertion for original (as-is, FSR) vs.\ LitMOF-corrected structures,
	restricted to the structures that also pass the \ce{CO2} $Q_\mathrm{st}$ DAC
	filter ($n = 287$). The original and corrected values are only weakly
	correlated (Pearson $r = 0.392$), in contrast to \ce{CO2}
	(\Cref{fig:widom_ff_co2}).}
	\label{fig:widom_ff_h2o}
\end{figure}

\begin{figure}[!ht]
	\centering
	\includegraphics[width=0.55\textwidth]{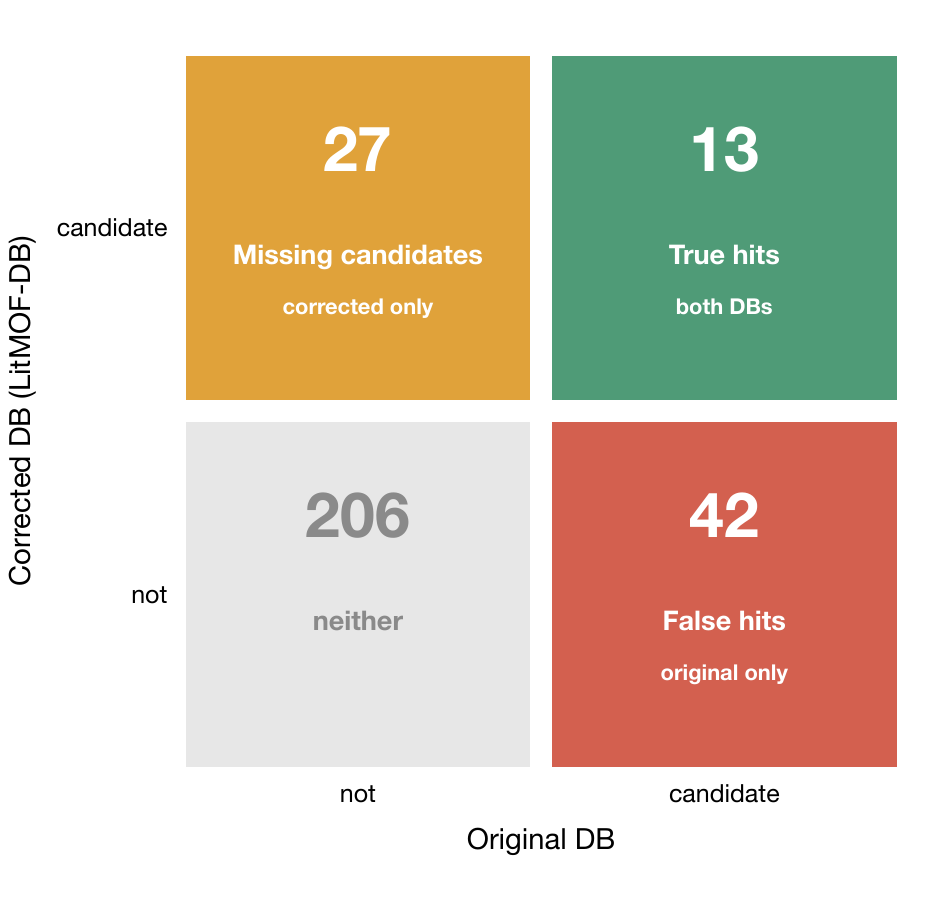}
	\caption{DAC candidate comparison between the original and corrected databases
	from classical force-field Widom screening. A candidate satisfies
	$|Q_\mathrm{st}(\ce{CO2})| > 40~\mathrm{kJ~mol^{-1}}$ and
	$S(\ce{CO2}/\ce{H2O}) > 1$. True hits are candidates in both databases,
	missing candidates only in the corrected database, false hits only in the
	original database, and the remaining structures fall in neither.}
	\label{fig:dac_contingency_ff}
\end{figure}

To confirm that the DAC screening results are not an artifact of the energy
method, we repeated the analysis with a classical force field and benchmarked
the dilute-limit Widom insertion against grand-canonical Monte Carlo (GCMC).
For all 4,256 MOFs retained after geometric filtering, we recomputed the
\ce{CO2} heat of adsorption with a classical force field (UFF\cite{rappe1992uff}
for the framework, TraPPE\cite{potoff2001vapor} for \ce{CO2}, and
TIP4P-Ew\cite{horn2004development} for \ce{H2O}) using Widom insertion in
RASPA2\cite{dubbeldam2016raspa}. Partial atomic charges were assigned with
PACMAN\cite{zhao2024pacman}.
The original structures again overestimate the strong-binding tail relative to
the corrected structures (\Cref{fig:widom_ff_co2}), and the resulting DAC
candidate comparison reproduces the trend seen with the MLIP, with more false
hits and missing candidates for the uncorrected database
(\Cref{fig:dac_contingency_ff}). We also performed \ce{H2O} Widom simulations on
the structures that pass the \ce{CO2} $Q_\mathrm{st} > 40~\mathrm{kJ~mol^{-1}}$
criterion (\Cref{fig:widom_ff_h2o}). Because \ce{H2O} binds relatively strongly,
the change between the original and corrected structures is more pronounced than
for \ce{CO2}, and the corrected $Q_\mathrm{st}(\ce{H2O})$ values correlate only
weakly with their original values (Pearson $r = 0.392$).

\begin{figure}[!ht]
	\centering
	\includegraphics[width=\textwidth]{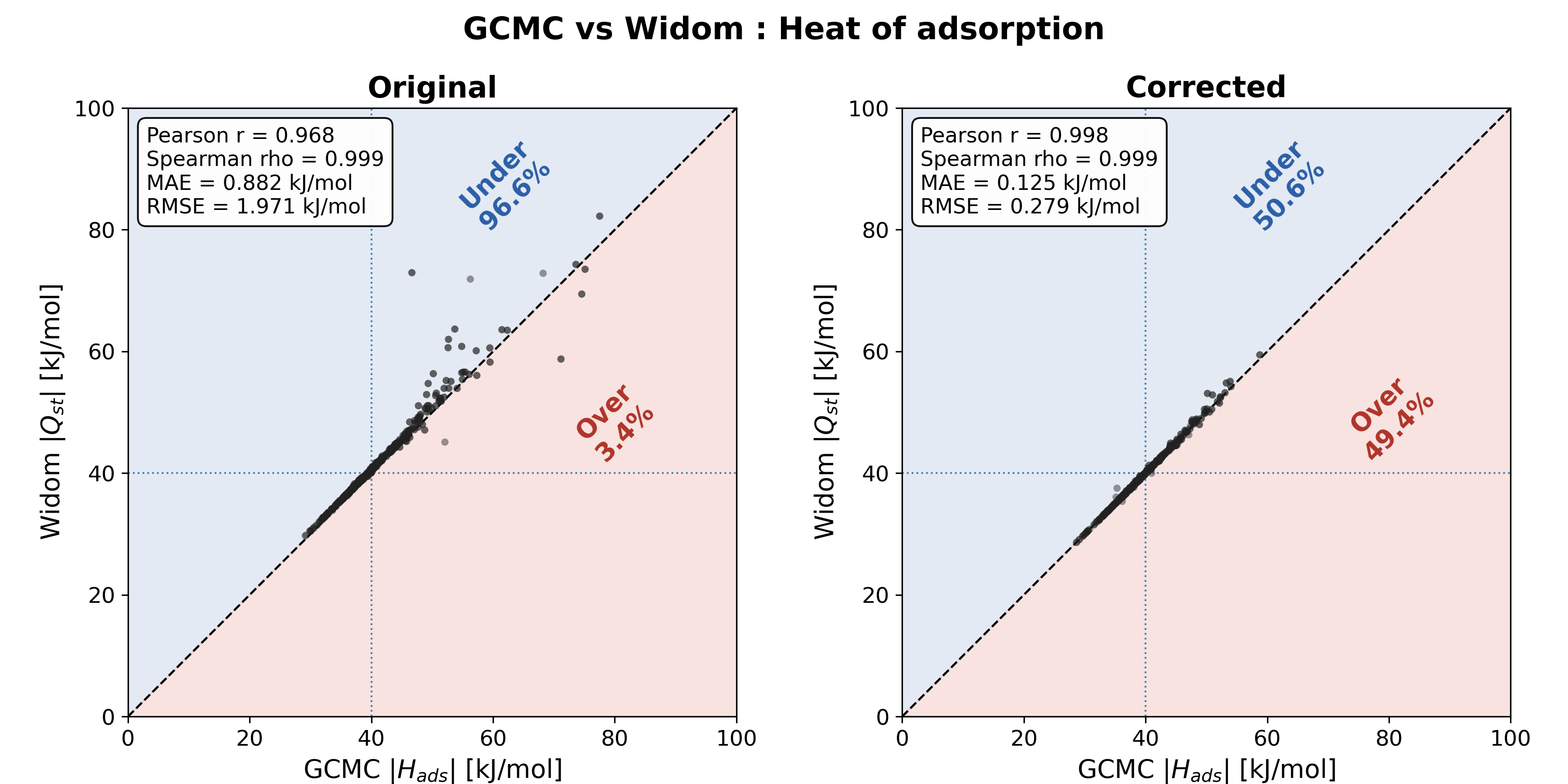}
	\caption{Comparison of $Q_\mathrm{st}(\ce{CO2})$ from classical force-field
	Widom insertion and GCMC for corrected and original structures.}
	\label{fig:gcmc_qst}
\end{figure}

\begin{figure}[!ht]
	\centering
	\includegraphics[width=\textwidth]{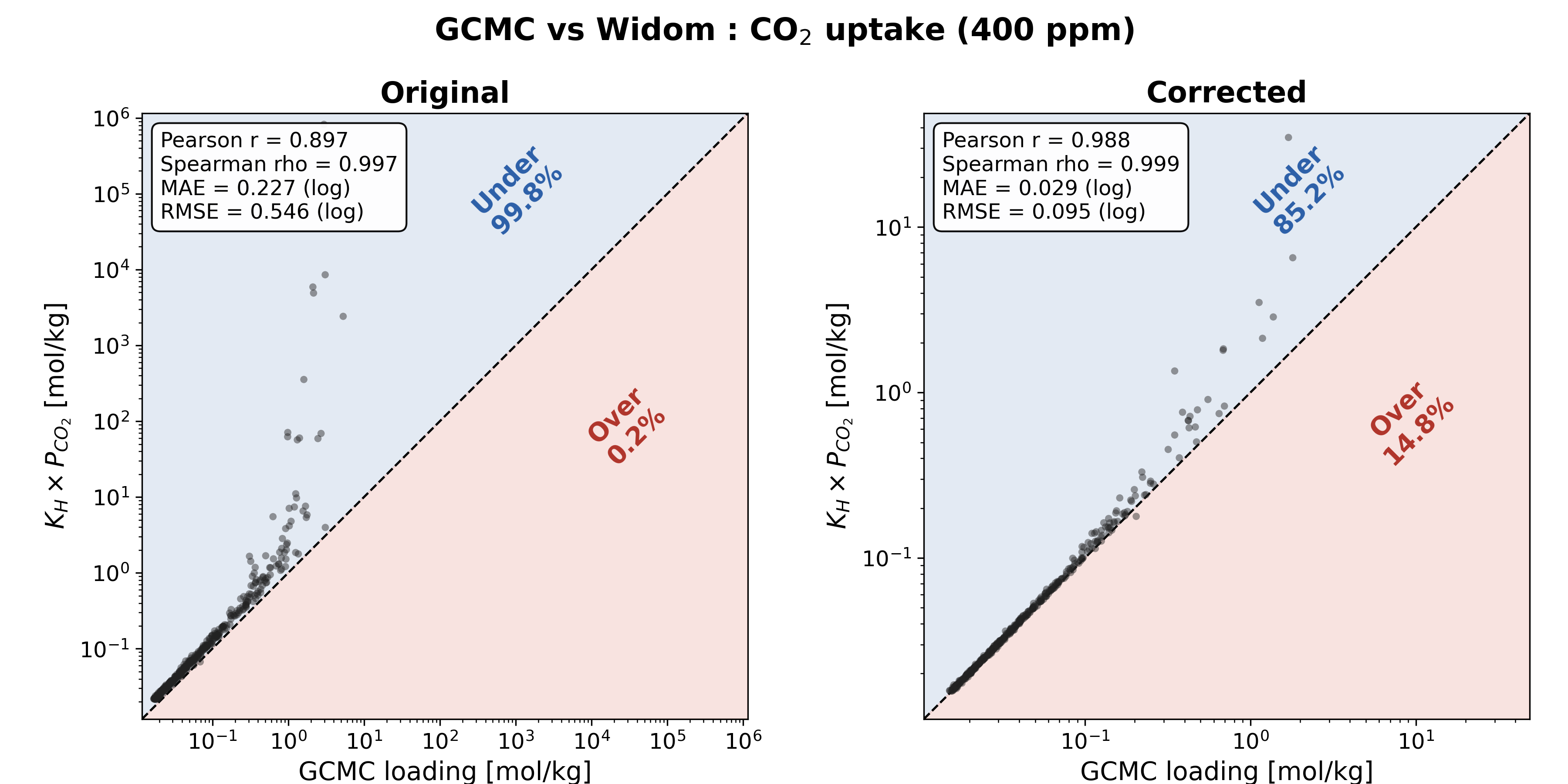}
	\caption{\ce{CO2} uptake at 400~ppm from classical force-field Widom insertion
	($K_\mathrm{H} \times 400~\mathrm{ppm}$) vs.\ GCMC. Pearson $r$ is computed on
	$\log_{10}$ uptake.}
	\label{fig:gcmc_uptake}
\end{figure}

To assess whether dilute-limit Widom insertion is a reliable surrogate for GCMC
at the DAC pressure of 400~ppm, we performed explicit GCMC simulations on the top
10\% of structures by $|Q_\mathrm{st}|$ in the union of the corrected and
original sets. For the corrected structures, Widom insertion and GCMC agree
closely (Pearson $r = 0.998$ for $Q_\mathrm{st}$ and $r = 0.988$ for
$\log_{10}$ uptake). A few of the highest-uptake candidates deviate at the top of
the range, but the corrected structures retain a largely monotonic Widom-GCMC
relationship (\Cref{fig:gcmc_qst} and \Cref{fig:gcmc_uptake}). For the original
structures, the agreement is substantially worse (Pearson $r = 0.968$ for
$Q_\mathrm{st}$ and $r = 0.897$ for $\log_{10}$ uptake). The structural errors in
these entries leave the molecular interactions in implausible configurations,
which propagate as severe artifacts into the computed values, so a substantial
number of samples show a large Widom-GCMC gap. Overall, the Widom results
reproduce the trend well at 400~ppm, whereas the uncorrected structures can
accumulate severe errors because of their underlying structural flaws.

\clearpage
\printbibliography

\end{refsection}

\end{document}